\documentclass[a4paper,fleqn,usenatbib]{mnras}

\usepackage{newtxtext,newtxmath}

\usepackage[T1]{fontenc}
\usepackage{ae,aecompl}

\usepackage{graphicx}	
\usepackage{amsmath}	
\usepackage{amssymb}	

\def\sarc{$^{\prime\prime}\!\!.$}

\newcommand{\OIII}{[O~{\sc iii}]}

\newcommand{\simgt}{\lower 2pt \hbox{$\, \buildrel {\scriptstyle >}\over {\scriptstyle\sim}\,$}}
\newcommand{\simlt}{\lower 2pt \hbox{$\, \buildrel {\scriptstyle <}\over {\scriptstyle\sim}\,$}}

\newcommand{\hs}{HS~0810$+$2554}

\title[Ultrafast and Molecular Outflows in \hs]{Linking the small scale relativistic winds and the large scale molecular outflows in the \MakeLowercase{{\sl z}} = 1.51 lensed quasar \hs}

\author[G. Chartas et al.]{
G. Chartas,$^{1}$\thanks{E-mail: chartasg@cofc.edu }
E. Davidson,$^{1}$
M. Brusa,$^{2,3}$
C. Vignali,$^{2,3}$ 
M. Cappi,$^{3}$ 
M. Dadina,$^{3}$
\newauthor{
G. Cresci,$^{4}$
R. Paladino,$^{5}$
G. Lanzuisi,$^{3}$
and A. Comastri$^{3}$}
\\
$^{1}$Department of Physics and Astronomy, College of Charleston, Charleston, SC, 29424, USA\\
$^{2}$Dipartimento di Fisica e Astronomia dell'Universit\`{a} degli Studi di Bologna, via P. Gobetti 93/2, 40129 Bologna, Italy\\
$^{3}$INAF, Osservatorio di Astrofisica e Scienza dello Spazio di Bologna, via P. Gobetti 93/3, 40129 Bologna, Italy\\
$^{4}$INAF, Osservatorio Astrofisico di Arcetri, Largo Enrico Fermi 5, I-50125 Firenze, Italy\\
$^{5}$INAF, Istituto di Radioastronomia, via Piero Gobetti 101, I-40129 Bologna, Italy
}

\date{Accepted XXX. Received YYY; in original form ZZZ}

\pubyear{2020}

\begin{document}
\label{firstpage}
\pagerange{\pageref{firstpage}--\pageref{lastpage}}
\maketitle

\begin{abstract}
We present Atacama Large Millimeter/submillimeter Array (ALMA) observations of the quadruply lensed $z~=~1.51$ quasar~\hs~which provide useful insight on the kinematics and morphology of the CO molecular gas and the $\sim$~2~mm continuum emission in the quasar host galaxy.
Lens modeling of the mm-continuum and the spectrally integrated CO(J=3$\rightarrow$2) images indicates that the source of the mm-continuum has an eccentricity of $e$~$\sim$~0.9 with a size of $\sim$ 1.6~kpc and the source of line emission has an eccentricity of $e$~$\sim$~0.7 with a size of $\sim$ 1~kpc.
The spatially integrated emission of the CO(J=2$\rightarrow$1) and CO(J=3$\rightarrow$2) lines shows a triple peak structure with the outer peaks separated by $\Delta{v_{21}}$~=~220~$\pm$~19 km~s$^{-1}$ and $\Delta{v_{32}}$~=~245~$\pm$~28~km~s$^{-1}$, respectively, suggesting the presence of rotating molecular CO line emitting gas.  Lensing inversion of the high spatial resolution images confirms the presence of rotation of the line emitting gas.
{Assuming a conversion factor of $\alpha_{\rm CO}$~=~0.8 $M_{\odot}$~(K~km~s$^{-1}$~pc$^{2}$)$^{-1}$ we find the molecular gas mass of \hs\ to be 
${ M }_{ Mol }$ ~=~(5.2 $\pm$ 1.5)/$\mu_{32}$~$\times$~10$^{10}$~$M_{\odot}$, where $\mu_{32}$ is the magnification of the CO(J=3$\rightarrow$2) emission.}
We report the {possible detection, at the 3.0$-$4.7$\sigma$ confidence level,} of shifted CO(J=3$\rightarrow$2) emission lines of high-velocity clumps of CO emission with velocities up to 1702~km~s$^{-1}$.
We find that the momentum boost of the large scale molecular wind is below the value predicted for an energy-conserving outflow given the momentum flux observed in the small scale ultrafast outflow.
\end{abstract}

\begin{keywords}
galaxies:active -- galaxies: high-redshift -- galaxies: emission lines -- ISM: jets and outflows -- quasars: general -- gravitational lensing: strong -- quasars: individual: \hs
\end{keywords}



\section{Introduction}

Powerful small scale X-ray absorbing winds have now been detected near the central supermassive black holes in several distant galaxies. These winds are believed to evolve into large scale outflows that regulate the evolution of the host galaxies. The mechanism by which these small scale winds transition into such important feedback mechanisms is not yet fully understood. Understanding the link between the small scale relativistic winds of Active Galactic Nuclei (AGN) and the associated large scale powerful molecular outflows in their host galaxies will thus provide valuable insight into the feedback process that regulates the growth of the supermassive black hole, the possible quenching of star formation and overall galaxy evolution. 
Theoretical models (e.g., \cite{2012MNRAS.425..605F}, \cite{2012ApJ...745L..34Z})  predict that AGN small scale winds initially collide with the interstellar medium gas and transfer their momentum to the host galaxy gas (momentum-conserving phase).
These models predict that at latter times the gas expands adiabatically in an energy-conserving mode with terminal velocities of about a few 1,000~km~s$^{-1}$.
Testing these models requires the comparison of the kinematics and the energetics of the small
and large scale outflows in the same objects and in AGN near the peak of AGN cosmic activity (i.e., at $z \sim 2-3)$.

Observations of ultraluminous infrared galaxies (e.g., \cite{2010A&A...518L..41F}, \cite{2010A&A...518L.155F}, \cite{2011ApJ...733L..16S}, \cite{2013ApJ...776...27V}, \cite{2014A&A...562A..21C}, \cite{2018A&A...612A..29B}, \cite{2019ApJ...887...69S}, \cite{2019A&A...628A.118B}, and \cite{2019MNRAS.489.1927S}) have revealed large-scale molecular outflows traced in OH and CO extending over kpc scales with velocities exceeding $\sim$ 1000 km s$^{-1}$ and with massive outflow rates (up to $\sim$1200 $M_{\odot}$ yr$^{-1}$).

The presence of both small and large scale energy-conserving outflows were recently discovered by \cite{2015Natur.519..436T}, 
in the $z$~=~0.189 ultraluminous infrared galaxy (ULIRG) IRAS F11119+3257 {(see also \cite{2017ApJ...843...18V} with updated estimates of the energy and momentum transfer in this object.)}, and by \cite{2015A&A...583A..99F}, in the $z = 0.04217$ ULIRG Mrk~231 (see also \cite{2018ApJ...867L..11L}, \cite{2019ApJ...871..156M}, and \cite{2019A&A...628A.118B}).
At $z>1$, the only object where both small and large scale outflows have been detected is the $z=3.912$ BAL quasar APM~08279+5255, showing molecular gas outflowing with maximum velocity of $v$~=~1,340~km~s$^{-1}$ \citep{2017A&A...608A..30F}.

Another promising quasar to test feedback models is the $z$~=~1.51 quadruply gravitationally lensed ULIRG \hs\
with a bolometric FIR luminosity (40$-$120 ${\mu}$m) of $L_{\rm FIR}$ = 10$^{13.5}$/$\mu_{\rm FIR}$~$L_{\odot}$, where $\mu_{\rm FIR}$ is the magnification factor in the FIR band \citep{2018MNRAS.476.5075S}.  
The total bolometric luminosity of \hs\ is $L_{\rm Bol}$ = 10$^{13.97}$/$\mu_{\rm UV}$~$L_{\odot}$, 
based on the monochromatic luminosities at 1450\AA \citep{2012MNRAS.422..478R}, where $\mu_{\rm UV}$$\sim$103 is the magnification factor in the UV band.

Our {\sl Chandra} and {\sl XMM-Newton} observations of \hs\ \citep{2016ApJ...824...53C}  indicate (at $>$99\% confidence) the presence of a highly ionized and relativistic outflow in this highly magnified object. The hydrogen column density of the X-ray outflowing absorber lies within the range $N_{\rm Habs}$~=~2.9--3.4~$\times$~10$^{23}$~cm$^{-2}$,
and the outflow velocity components lie within the range  $v_{\rm out}$~=~0.1$-$0.4 $c$.
The mass-outflow rate of the X-ray absorbing material of \hs\ is found to lie in the range of $\dot{M}$~=~1.5$-$3.4~M$_{\odot}~yr^{-1}$ and is comparable to the accretion rate of $\sim$ 1 M$_{\odot}~yr^{-1}$.
UV spectroscopic observations with VLT/UVES \citep{2016ApJ...824...53C}  indicate that the UV absorbing material of \hs\ is outflowing at $v_{\rm UV}$$\sim$19,400~km~s$^{-1}$. VLA observations of \hs\  \citep{2015MNRAS.454..287J} at 8.4~GHz have revealed radio emission in this radio-quiet object.

More recently, \cite{2019MNRAS.485.3009H} using e-MERLIN and European VLBI Network observations have 
identified jet activity in \hs.  Specifically, their source reconstruction of the radio and HST data of \hs\
shows two jet components linearly aligned on opposing sides of the optical quasar core.
\cite{2018MNRAS.476.5075S} found that \hs\ falls on the radio$-$FIR correlation, even though, the observations of 
Hartley et al. show that the radio emission of \hs\ arises predominately from a jetted quasar. Consequently they suggest that
the radio$-$FIR correlation cannot always be used to rule out AGN activity in favor of star formation activity.
We also note that the SED template fit to \hs\ indicates that the AGN contribution in this object is significant \citep{2010MNRAS.406..720R} and a strong starburst in this object is unlikely. Specifically, the ratio of the torus to optical luminosity of \hs\ is found to be $L_{\rm torus}$/$L_{\rm opt}$ $\sim$ 0.5, which can be interpreted as a dust covering factor.

We expect that the small scale ultrafast outflows in \hs\ may be driving a larger scale outflow of molecular gas in the host galaxy.  
To investigate the possible presence of a large scale outflow we obtained  ALMA observations of \hs\ in cycle 5. 
The main goals of our ALMA observations were to (a) spatially resolve the continuum mm emission emitted 
in the host galaxy of \hs\ and infer its physical extent, 
and (b) detect the molecular gas in CO(J=3$\rightarrow$2) and CO(J=2$\rightarrow$1), measure the total molecular CO mass and its extent, and 
search for a possible outflow.

In Section~2  we present the ALMA observations and data reduction of \hs. 
In Section~3 we present the analysis of the mm continuum emission.
In Section~4 we present the analysis of the  CO(J=3$\rightarrow$2)  emission performed performed with the ALMA extended configuration. In Section~5 we present the analysis of the  CO(J=3$\rightarrow$2) and CO(J=2$\rightarrow$1) emission performed with the ALMA compact configuration. 
In Section~6 we present the results of our lensing analysis to explain the offset between the optical and mm image positions and our results from modeling the
spatially integrated spectra of the CO(J=3$\rightarrow$2) and CO(J=2$\rightarrow$1) emission lines.
In Section~7 we present the possible detection of outflows of CO(J=3$\rightarrow$2) emitting molecular gas.
Finally, in Section~8 we present a discussion of our results and conclusions.
Throughout this paper we adopt a flat $\Lambda$ cosmology with 
$H_{0}$ = 68~km~s$^{-1}$~Mpc$^{-1}$ $\Omega_{\rm \Lambda}$ = 0.69, and  $\Omega_{\rm M}$ = 0.31 \citep{2016A&A...594A..13P}.

\section{ALMA Observations and Data Reduction} \label{sec:data}
\hs~was observed with ALMA during cycle 5 (project 2017.1.01368) in Band 3 (100~GHz) and Band 4 (140~GHz), where the redshifted  CO(J=2$\rightarrow$1) and  CO(J=3$\rightarrow$2) transitions are visible, respectively.
Table \ref{tab:obslog} summarizes the observation dates, source integration science times, and spectral configuration. 

The Band 3 observations have been taken in one session in the configuration C43-5, with a maximum baseline of 1.4 km, providing a spatial resolution of $\sim$ 1\arcsec. The spectral configuration used provided 1920 channels with a spectral resolution of 3.2 km~s$^{-1}$.
The Band 4 observations have been taken in two sessions: the first one with the extended configuration C43-7, providing a spatial resolution of $\sim$ 0.1 \arcsec, in TDM (Time Division Mode) spectral configuration, with 128 channels, {34 km~s$^{-1}$} wide.
The second session has been taken with a more compact configuration, C43-3, providing a resolution of $\sim$ 1.2\arcsec, and a higher spectral resolution, 1920 channels, 2.1 km~s$^{-1}$ wide.

Each ALMA data set has been calibrated using the ALMA pipeline (version Pipeline-CASA51-P2-B).
The calibrated data have been imaged and analyzed with the CASA software (version 5.1.1-5).
The images of continuum emission in both bands have been obtained running the CASA task {\sl tclean} in the multi-frequency synthesis (mfs) mode on the visually selected line-free channels of the observed spectral ranges. To obtain the maximum sensitivity we used the Briggs weighting scheme with the robust parameter set to 2 (natural weighting). The achieved rms in the continuum images are 16 $\mu$Jy~beam$^{-1}$ in Band 3, 13 $\mu$Jy~beam$^{-1}$ and 29 $\mu$Jy~beam$^{-1}$ in the Band 4 high and low spatial resolution images, respectively.

Prior to creating the CO line cubes, the continuum has been subtracted from the datasets using the CASA  task {\sl uvcontsub}. A linear fit of the continuum in the visually selected line-free channels of the spectral windows has been calculated in the visibility plane and subtracted from the full spectral range.
The datacube for the high spatial resolution image in band 4 has been obtained using the original spectral resolution of the data (34 km~s$^{-1}$).
The datacubes for the band 3 and band 4 low spatial resolution images have been obtained by binning the spectra by factors of 8 and 4, respectively
(reaching velocity resolutions of 25.6 km~s$^{-1}$ and 8.4 km~s$^{-1}$, respectively).

The natural weighting scheme, that was used for the continuum images, has also been applied to the CO line cubes to enhance the sensitivity.
{The rms values achieved in the line datacubes have been measured in line-free channels and are found to be
$\sim$ 150 $\mu$Jy~beam$^{-1}$ for a 34.0 km~s$^{-1}$ channel width in Band 4 (extended configuration),
$\sim$ 390 $\mu$Jy~beam$^{-1}$ for a 25.6 km~s$^{-1}$ channel width in Band 3 (compact configuration), and
$\sim$ 750 $\mu$Jy~beam$^{-1}$ for a 34.0 km~s$^{-1}$ channel width in Band 4 (compact configuration).}
We finally obtained the integrated intensity images (moment 0 maps) of each datacube, using the task {\sl immoments}.

\section{Analysis of the mm Continuum Emission of \hs.} \label{sec:continuum}
\begin{figure}
\includegraphics[width=\columnwidth]{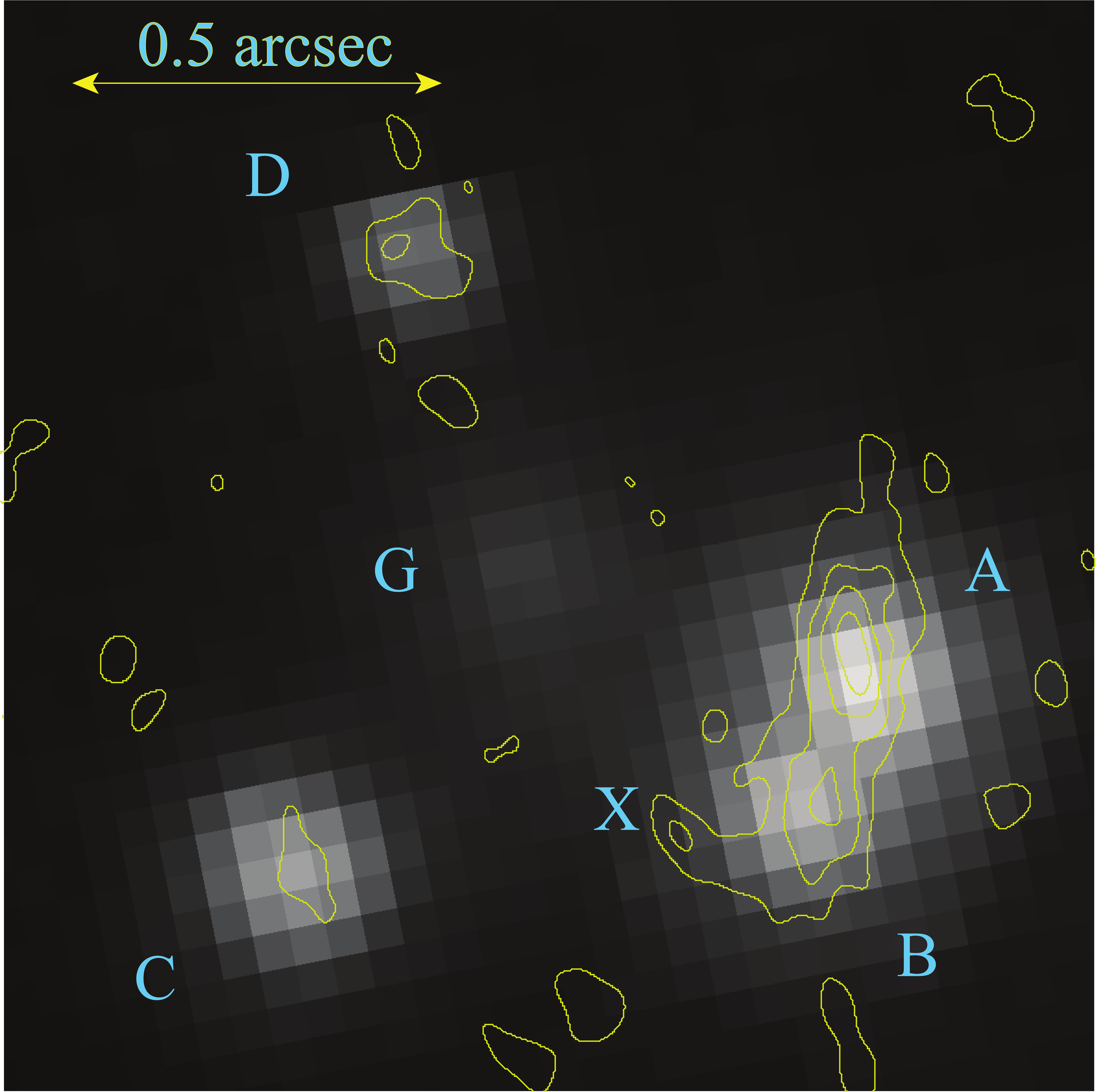}
\caption{The {\sl HST} ACS F555W image of the lensed $z$=1.51 quasar \hs\
with the {\sl ALMA} band 4 continuum contours overlaid. The offset between the
optical and the mm continuum images can be explained by lensing
effects and different emission locations and sizes of the optical and mm-emission regions.
The mm-continuum image shows a new feature labelled X near image B that is not present in the optical.
G is the lens galaxy. The contours represent the $\sim$2~mm continuum  [-3, -2, 2, 3, 4, 5]$\sigma$ levels, where 1$\sigma$ = 12.7 $\mu$Jy~beam$^{-1}$.
\label{fig:hst_alma}}
\end{figure}

A mm-continuum image was created from the high spatial resolution dataset using multifrequency synthesis with the channels
containing the CO(J=3$\rightarrow$2) line removed. 
In Figure \ref{fig:hst_alma} we show the HST ACS F555W image of \hs\ with the ALMA band 4 continuum contours overlaid.
{The HST and the ALMA band 4 continuum images were aligned relative to the isolated image D.}
The mm-continuum lensed images of the source A and B are detected as extended emission at a combined significance level of  $\sim$10$\sigma$, however, images C ($\sim$ 1.6$\sigma$)  and D ($\sim$ 2.3$\sigma$) barely rise above the noise level (see Table \ref{tab:pos}).

In Table \ref{tab:pos} we list the relative positions of the HST image positions with respect to image A taken from the CfA-Arizona Space Telescope LEns Survey (CASTLES) of gravitational lenses,
the relative positions of the ALMA continuum images with respect to A and the mm-continuum flux densities of the images.  The peaks of the mm-continuum images  A and B are found to be offset at the $> 10\sigma$ level from the optical centroids when aligning the optical and mm-continuum maps to the isolated image D. 
This offset is investigated in the lens modeling of \hs\ presented in section 6.

The mm-continuum emission also shows a new feature labelled X near image B that is not present in the optical image. Component X also does not align with any of the radio components resolved in the
e-MERLIN observations at 4.32~GHz and 5.12~GHz of \hs\ \citep{2019MNRAS.485.3009H}.

The flux density of the continuum of the combined images at a mean frequency of 143~GHz is estimated to be (401 $\pm$ 50)/$\mu_{\rm cont}$ $\mu$Jy,
where $\mu_{\rm cont}$ = 7 $\pm$ 3 is the lensing magnification of the mm-continuum emitting region estimated in section 6.

The mm-continuum flux density of \hs\ was also estimated from the low spatial resolution observations. 
Specifically, for the August and March 2018 observations performed in Band 4, the mm-continuum image was created using multifrequency synthesis
with the channels containing the CO(J=3$\rightarrow$2) line removed. For the January 2018 observation performed in Band 3 
the channels containing the CO(J=2$\rightarrow$1) line were removed.
To estimate the flux density of the continuum we used the CASA tool 2D Fit with the fitting region set to a circle of radius of 2\sarc5  centered on the source.
We find the flux densities of the continuum at mean frequencies of 143~GHz and  97~GHz  to be 
(400 $\pm$ 38 $\mu$Jy)/$\mu_{\rm cont}$ and (197 $\pm$ 49)/$\mu_{\rm cont}$ $\mu$Jy, respectively. 
As we predicted, the observed ALMA flux densities are found to be just below the 
3$\sigma$ upper limit of $\sim$1 mJy inferred from the non-detection of \hs\ by 
CARMA \citep{2011ApJ...730..108R}.

\section{The Einstein Ring of the CO(J=3$\rightarrow$2) Emission is resolved with the ALMA extended configuration.}
\begin{figure}
\includegraphics[width=\columnwidth]{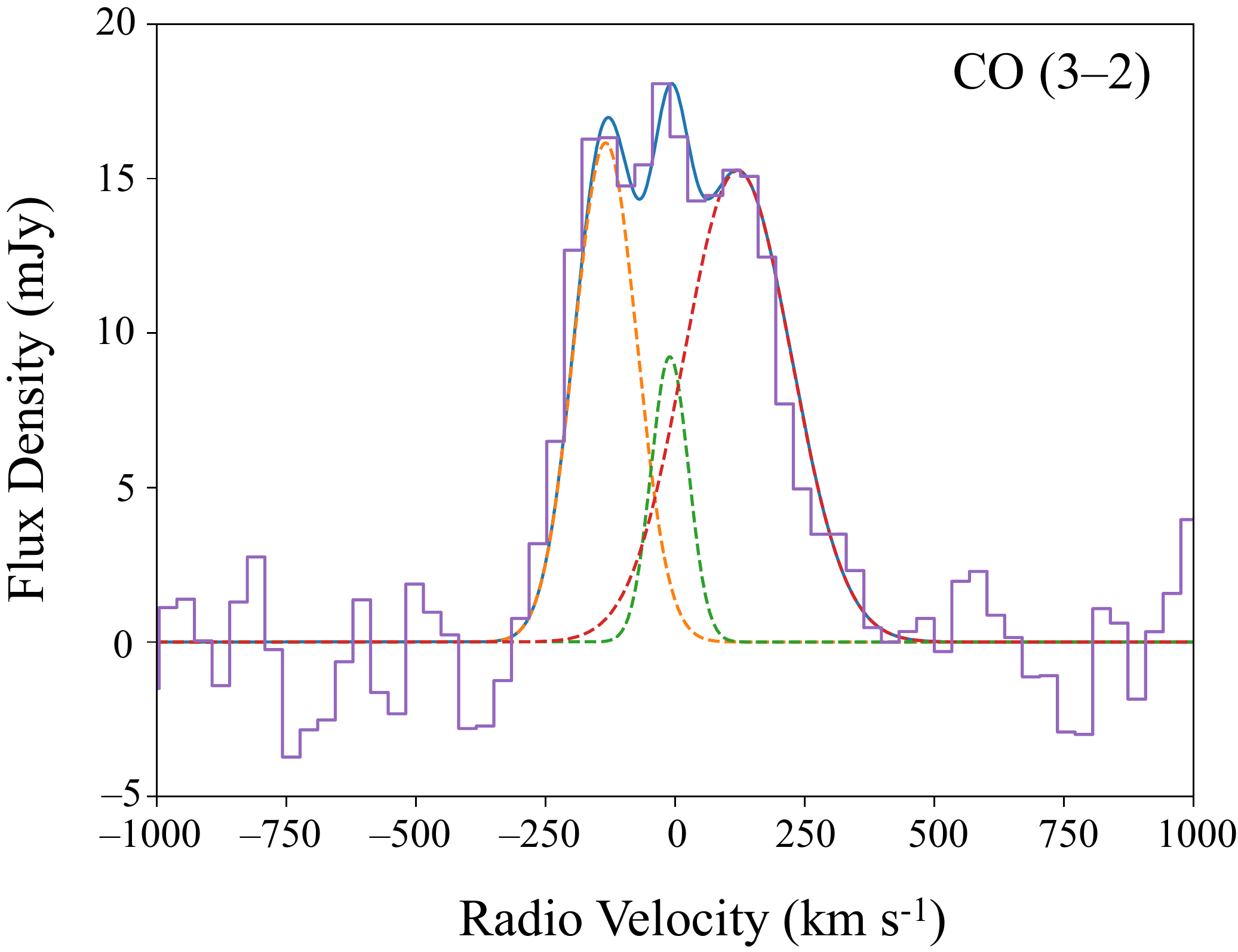}
\caption{The ALMA CO(J=3$\rightarrow$2) continuum-subtracted spectrum of \hs\  obtained with the ALMA extended configuration.
The spectral line is fit with a model that consists of 3 Gaussians. The velocity shifts can be explained with the presence of a rotating molecular disk.
We have assumed a systemic redshift of $z = 1.50849 \pm 0.00002 $ based on the redshift of the \OIII\ line detected in the 
Spectrograph for INtegral Field Observations in the Near Infrared (SINFONI) spectrum of \hs\ (Cresci et al., in prep).
\label{fig:alma_co32_linespec}}
\end{figure}

The image of the CO(J=3$\rightarrow$2) line emission was produced
from the continuum subtracted data, 
in the frequency range 137.714~GHz to 137.980~GHz.
Also in this case, we used natural weighting to maximize the S/N (Briggs parameter R = 2) for our image of the line emission. 
A spatially integrated spectrum was extracted from a circular region centered on \hs\ with
a radius of 0\sarc7. The resulting spectrum shown in Figure \ref{fig:alma_co32_linespec} 
shows a prominent (S/N $\sim $9) emission line with a clear asymmetric triple peaked structure. 
The spectrum was initially fit with a model consisting of a single Gaussian.  This fit is not acceptable in a statistical sense with
 ${\chi^2}/\nu = 189/16$, where $\nu$ are the degrees of freedom. 
We next use a model consisting of three Gaussians. The fit with three Gaussians
results in an acceptable fit with  ${\chi^2}/\nu$ = 11.4/10.
 The best-fit parameters of the centroid velocities and full-width-half-maxima (FWHM) were found to be 
($v_1$, FWHM$_1$) = ($-$134 $\pm$ 10~km~s$^{-1}$, 142  $\pm$ 22 ~km~s$^{-1}$),  
($v_2$, FWHM$_2$) = ($-$11 $\pm$ 11~km~s$^{-1}$, 85  $\pm$ 31 ~km~s$^{-1}$), 
and ($v_3$, FWHM$_3$ )= (120 $\pm$ 21~km~s$^{-1}$, 244 $ \pm$ 39 ~km~s$^{-1}$).
{The dominant blue and red peaks of the CO(J=3$\rightarrow$2) line profile are suggestive of rotation. If this is the case,
then the velocity separation between the outer peaks corresponds to a rotational velocity of $\sim$ 250 km~s$^{-1}$ (without a correction for inclination).
The CO line profiles of lensed AGN will in general be distorted by differential magnification across the molecular disk (e.g., \cite{2015MNRAS.453L..26R}, \cite{2018A&A...613A..34P}, \cite{2017ApJ...836..180L}).
In section 7, as part of our lens modeling we derive the moment 1 velocity map of the CO(J=3$\rightarrow$2) line emission of \hs\ calculated in the source plane. This velocity map clearly shows disk rotation with a velocity range that is consistent with the CO(J=3$\rightarrow$2) line profile.}
We have assumed a systemic redshift of 
$z = 1.50849 \pm 0.00002$ based on the redshift of the \OIII\ line detected in the 
Spectrograph for INtegral Field Observations in the Near Infrared (SINFONI) spectrum of \hs\ (Cresci et al., in prep)
The integrated flux density of the CO(J=3$\rightarrow$2) emission line is (6.8~$\pm$~0.8~Jy~km~s$^{-1}$)/$\mu_{32}$,
where $\mu_{32}$ is the lensing magnification of the CO(J=3$\rightarrow$2) emitting region estimated in section 6.
{The integrated flux densities of the three best-fit gaussian lines that make up the CO(J=3$\rightarrow$2) line (see Figure 2) are $S_{1}$~=~2.37~$\pm$~0.56~Jy~km~s$^{-1}$, $S_2$ = 0.84 $\pm$ 0.16 Jy km~s$^{-1}$, and $S_3$ = 3.55 $\pm$ 0.40 Jy km~s$^{-1}$.
The rms noise will mostly affect the estimated integrated flux density of the weakest central line component. However, as we show in section 5,	the central line component is also
detected in the spectra of the CO(J=3$\rightarrow$2) and CO(J=2$\rightarrow$1) lines obtained with the ALMA compact configuration.}

\begin{figure}
\includegraphics[width=\columnwidth]{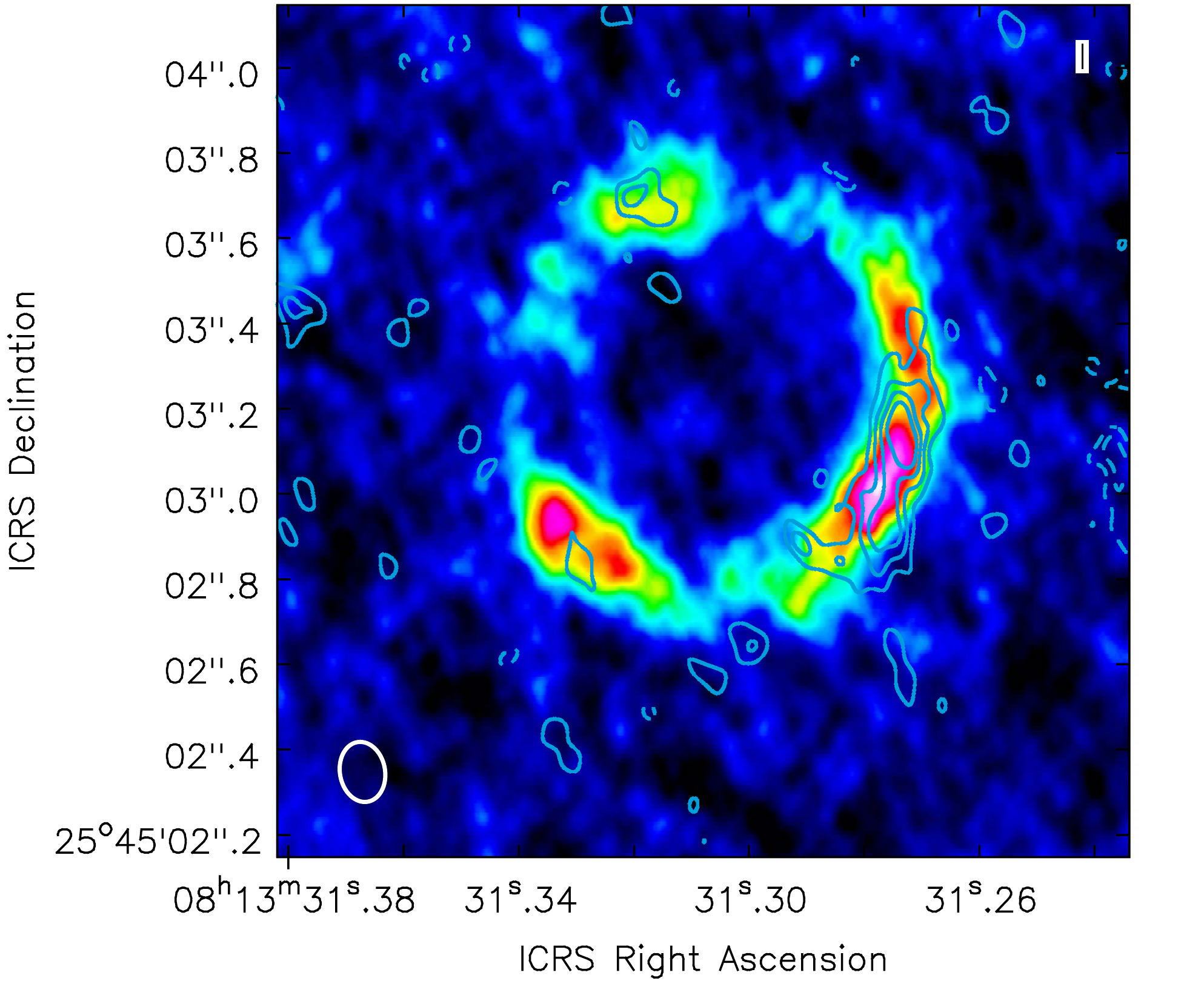}
\caption{The ALMA CO(J=3$\rightarrow$2) moment 0 line emission of \hs\ with the 
2 mm ALMA continuum overlaid as contours.
The ALMA synthetic beam size is 0.1$\arcsec$ $\times$ 0.06$\arcsec$ and is indicated with the solid white ellipse in the lower left corner. 
The CO(J=3$\rightarrow$2) is clearly more extended along an Einstein ring than the 2mm continuum emission.
The contours represent the $\sim$2~mm continuum [-3, -2, 2, 3, 4, 5]$\sigma$ levels, where 1$\sigma$ = 12.7 $\mu$Jy~beam$^{-1}$. Solid and dashed contours represent positive and negative values.
\label{fig:alma_cont_line}}
\end{figure}

\begin{figure*}
\includegraphics[width=17.5cm]{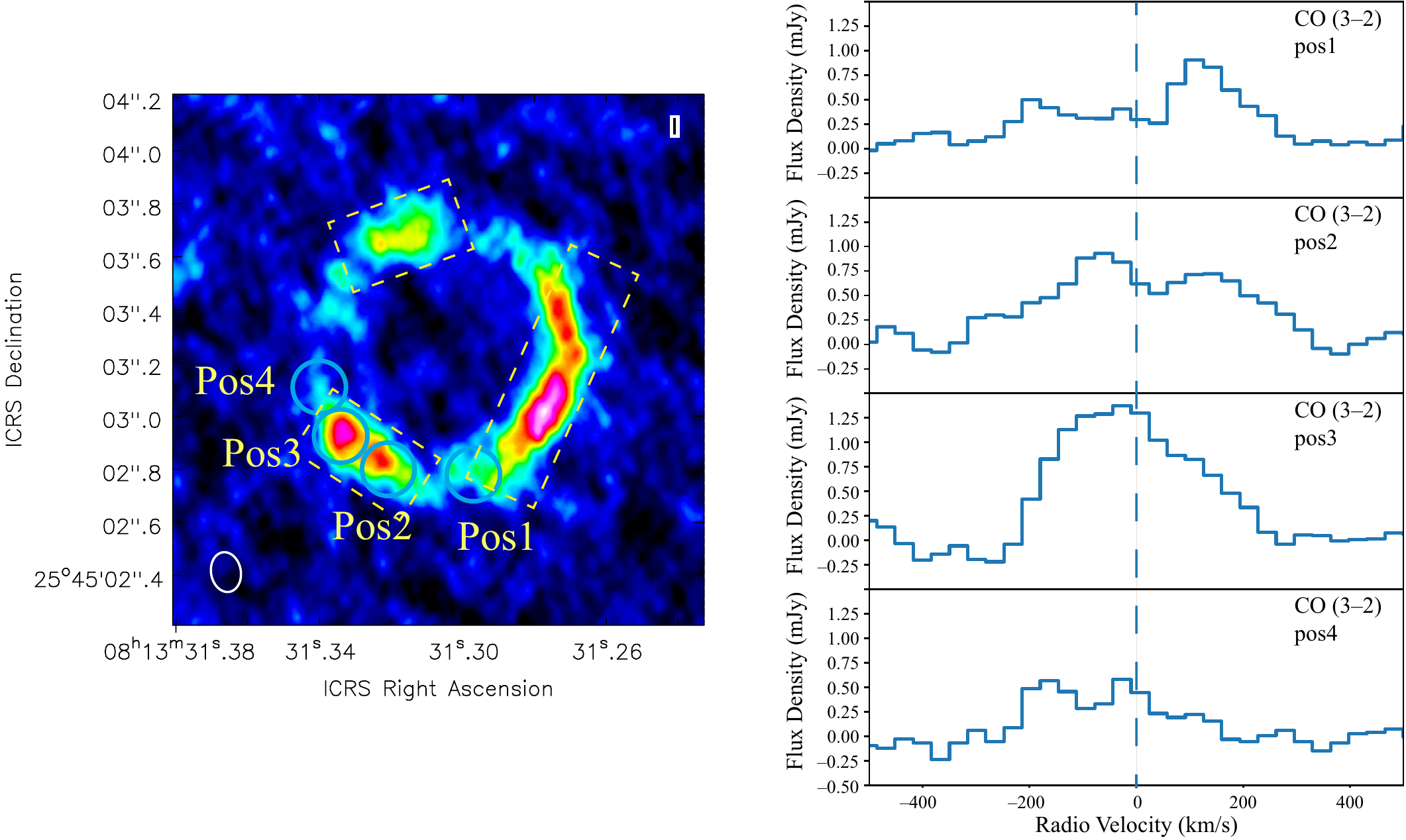}
\caption{(Left) The image of \hs\ spectrally integrated over the 
CO(J=3$\rightarrow$2) line with solid circles representing the extraction regions used to produce the spectra on the right. (right) CO(J=3$\rightarrow$2) spectra extracted from four different locations on the Einstein Ring near image C. The observed Doppler shift is consistent with a rotating CO(J=3$\rightarrow$2) emission region. The dashed rectangular regions represent the three slices used to produce the position velocity diagrams shown in Figure \ref{fig:pv_diagram}
\label{fig:alma_co32_doppler}}
\end{figure*}

\begin{figure}
\includegraphics[width=\columnwidth]{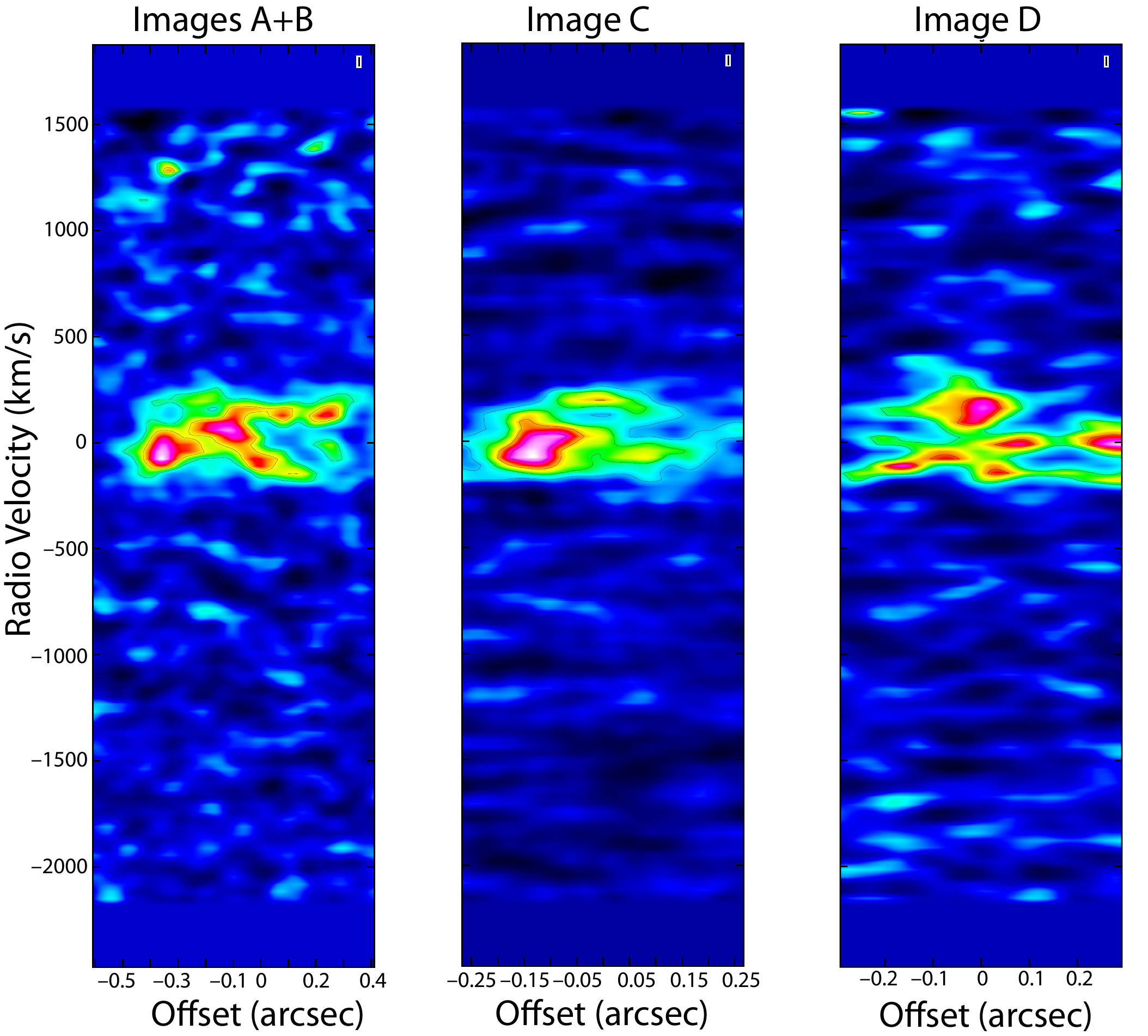}
\caption{
Position velocity diagrams for images A+B, C, and D of the high spatial resolution CO(J=3$\rightarrow$2) image cube of \hs.
The position velocity diagrams  are computed over three slices in the direction plane centered on images A+B, C and D (see Figure \ref{fig:alma_co32_doppler}).
The start and endpoint positions of the slices correspond to negative and positive offsets following the 
Einstein ring along the clockwise direction. 
Contours overlaid correspond to [2, 3, 5, 7]$\sigma$ for images A+B, 
 [2, 3, 4, 5]$\sigma$ for images C, and  [2, 3, 4]$\sigma$ for images D, where 1$\sigma$~=~140~$\mu$Jy~beam$^{-1}$.
Redshifted clumps with radio velocities of $\sim$ 1300 km~s$^{-1}$are detected in the slice over images A+B (also see clump 1 in Figure \ref{fig:alma_clumps}). 
\label{fig:pv_diagram}}
\end{figure}

\begin{figure}
\includegraphics[width=\columnwidth]{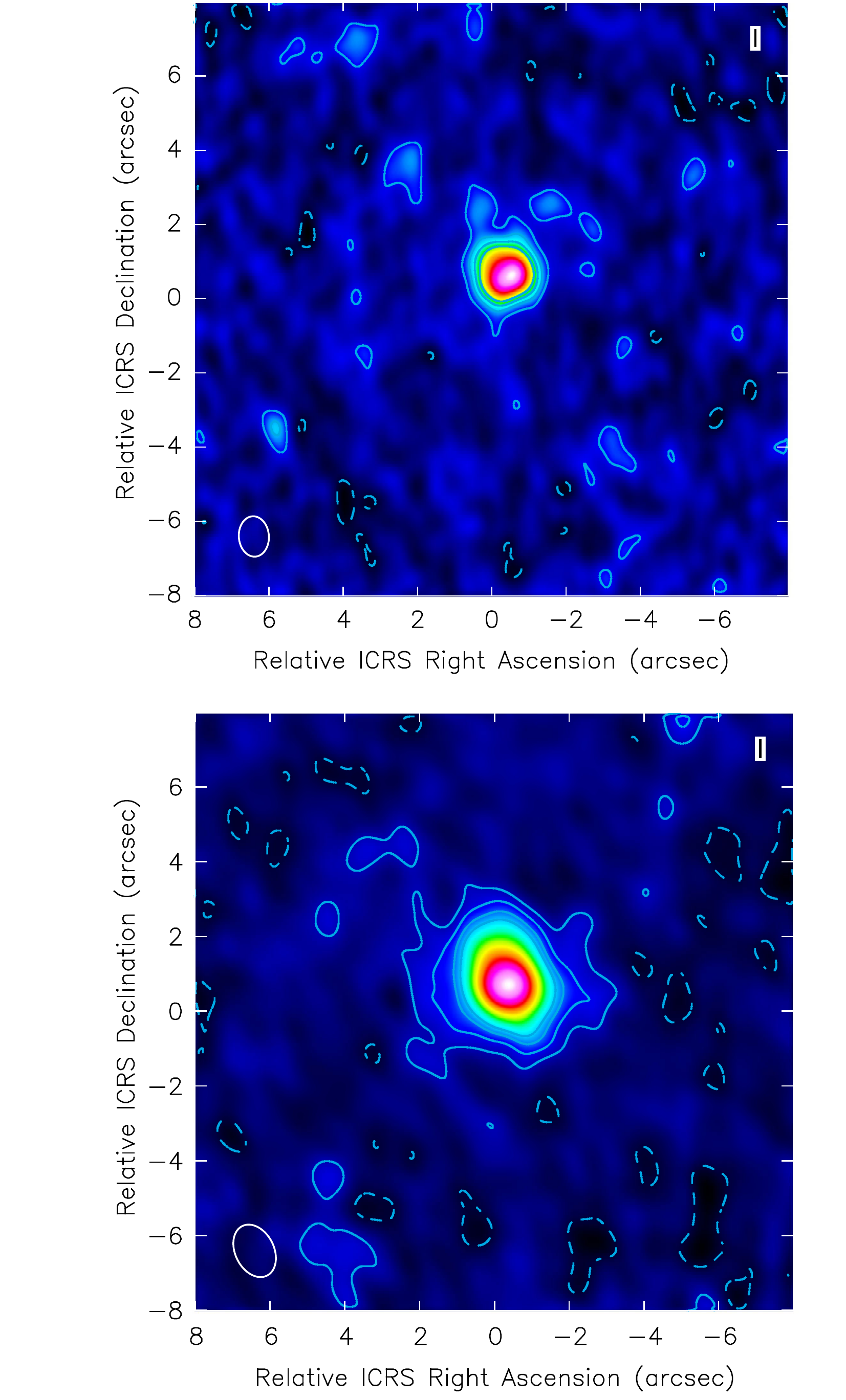}
\caption{ (Top) The ALMA CO(J=2$\rightarrow$1) moment 0 line emission of \hs . Contours correspond to [-2, 2, 4, 6, 8]$\sigma$, where 1$\sigma$ = 0.05~Jy~beam$^{-1}$~km~s$^{-1}$.
(bottom) The ALMA CO(J=3$\rightarrow$2) moment 0 line emission of \hs\  obtained with the ALMA compact configuration.
Contours correspond to [-2, 2, 4, 6, 8]$\sigma$, where 1$\sigma$ = 0.08~Jy~beam$^{-1}$~km~s$^{-1}$.
\label{fig:alma_m0_lowres}}
\end{figure}

\begin{figure}
\includegraphics[width=\columnwidth]{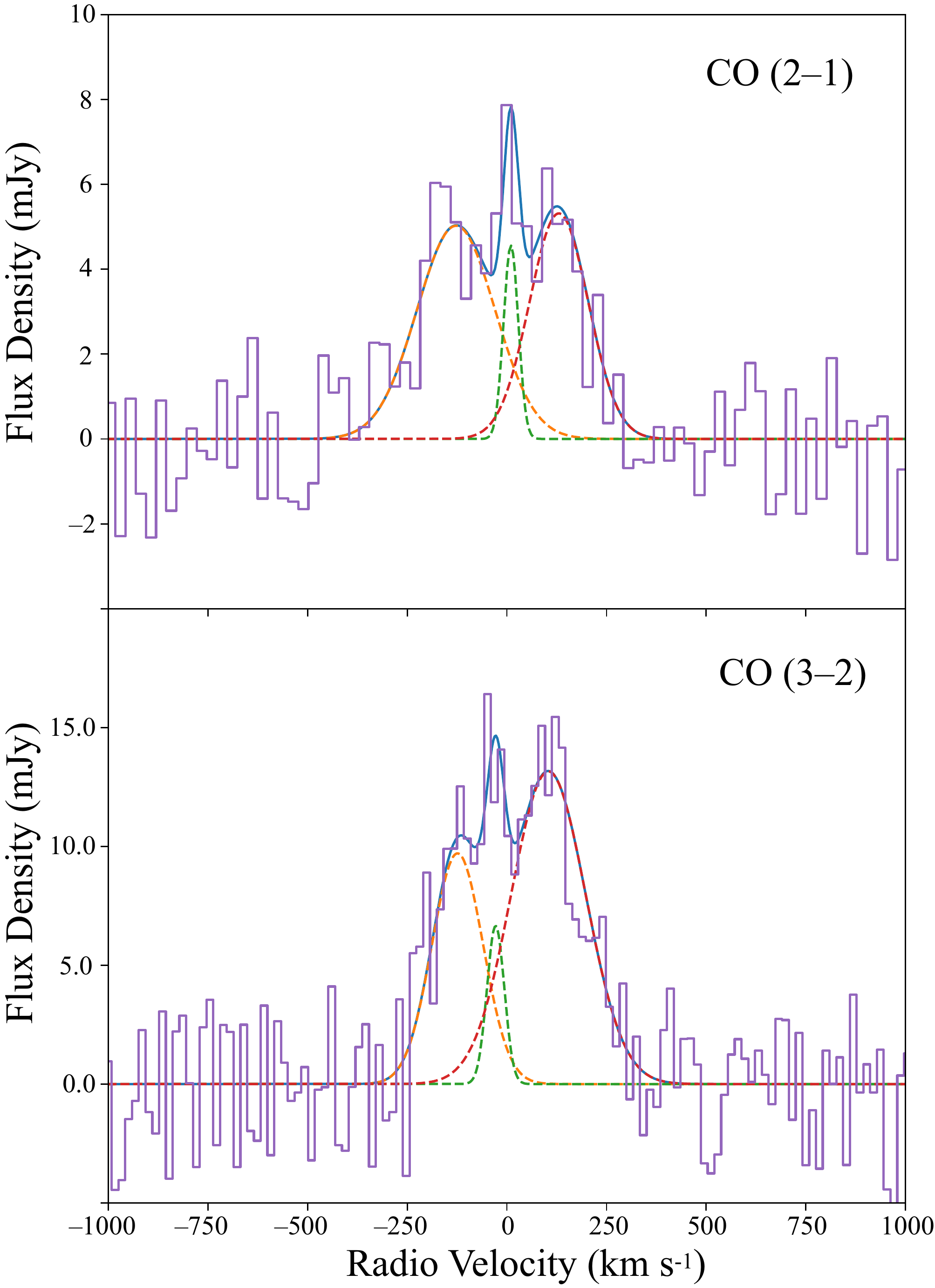}
\caption{The ALMA CO(J=3$\rightarrow$2) and CO(J=2$\rightarrow$1) continuum-subtracted spectra of \hs\  obtained with the ALMA compact configuration. 
The spectral lines are fit with a model that consists of 3 Gaussians. The velocity shifts can be explained with the presence of a rotating molecular disk and are 
consistent with the best-fit values derived from the CO(J=3$\rightarrow$2) spectrum obtained with the ALMA extended configuration.
\label{fig:alma_co32_co21_lines}}
\end{figure}

We produced spectrally integrated maps (moment 0 images) of the line and
continuum emission components by integrating over the entire spectrum of these components. 
In Figure \ref{fig:alma_cont_line} we show a raster image of the moment 0 map of the CO(J=3$\rightarrow$2) line
with overlaid contours.
The moment 0 map of the CO(J=3$\rightarrow$2)  line shows that the line emission forms a
partial Einstein ring in addition to emission centered around the 4 images A, B, C, and D. 
A clear offset is present between the peaks of images A and B in the lensed mm-continuum  and CO(J=3$\rightarrow$2) line emission of \hs\
in the sense that the line connecting the mm-continuum peaks of images A and B is rotated counter-clockwise 
with respect to the tangent to the CO(J=3$\rightarrow$2) Einstein ring at the midpoint between these images.
This offset is similar to the offset shown in Figure \ref{fig:hst_alma} between the peaks of images A and B in the optical and mm-continuum emission.

Gravitational lensing of \hs\ stretches the extended CO(J=3$\rightarrow$2) source emission in each image along the Einstein radius. Spectra extracted along the Einstein radius show a Doppler shift of the centroid of the CO(J=3$\rightarrow$2) line as the spectral extraction region is moved along the Einstein radius. We illustrate the disk rotation in Figure \ref{fig:alma_co32_doppler}  by showing the CO(J=3$\rightarrow$2) spectra extracted from four different locations on the Einstein Ring near the brightest isolated image C. The observed Doppler shift is consistent with a rotating CO(J=3$\rightarrow$2) emission region.

In Figure \ref{fig:pv_diagram} we show the position velocity diagrams for images A+B, C, and D of the high spatial resolution CO(J=3$\rightarrow$2) image cube of \hs.
Note that images A and C have positive parity and images B and D have negative parity (mirrored images of the source). 
The position velocity diagrams were created using the CASA task {\sl impv} and are computed over three slices in the direction plane centered on images A+B, C and D. The slices are shown in Figure \ref{fig:alma_co32_doppler}.
A single slice is used to construct the position velocity diagram of images A+B since the elongated images A and B overlap.
The widths of the slices are 0\sarc25. The offsets are with respect to the center of the slices.
The start and endpoint positions of the slices correspond to negative and positive offsets following the 
Einstein ring along the clockwise direction. 

\section{Analysis of the  CO(J=3$\rightarrow$2) and CO(J=2$\rightarrow$1) emission lines obtained with the ALMA compact configuration.} \label{sec:co32_21}

The images of the CO(J=3$\rightarrow$2) and CO(J=2$\rightarrow$1) line emission of \hs\ observed with the ALMA compact configuration were produced by subtracting the continuum spectral component from the spectral channels containing the CO(J=3$\rightarrow$2) line in the frequency range 137.7~GHz to 138.0~GHz and CO(J=2$\rightarrow$1) line in the frequency range 91.8~GHz to 92.0~GHz.
In Figure  \ref{fig:alma_m0_lowres} we show the moment 0 images of the CO(J=3$\rightarrow$2) and CO(J=2$\rightarrow$1) line emissions overlaid with [-2, 2, 4, 6, 8]$\sigma$ contours on the line emissions, where $\sigma$  is the rms sensitivity.

We used natural weighting for our image of the line emission. 
Spatially integrated spectra of the CO(J=3$\rightarrow$2) and CO(J=2$\rightarrow$1) line emission were extracted from circular regions centered on \hs\ with radii of 3~\arcsec. 
The resulting spectra shown in Figure \ref{fig:alma_co32_co21_lines} show multiple peaks. 

The CO(J=3$\rightarrow$2)  spectrum was initially fit with a model consisting of a single Gaussian.  This fit is not acceptable in a statistical sense with
 ${\chi^2}/\nu = 49/23$, where $\nu$ are the degrees of freedom. 
We next use a model consisting of three Gaussians. The fit with three Gaussians
results in an acceptable fit with  ${\chi^2}/\nu = 18.5/17$.The best-fit parameters for the centroid velocities and FWHM of the CO(J=3$\rightarrow$2) line were found to be ($v_1$, FWHM$_1$) = ($-$124 $\pm$ 22~km~s$^{-1}$, 151  $\pm$ 45 ~km~s$^{-1}$),  
($v_2$, FWHM$_2$) = ($-$28 $\pm$ 9~km~s$^{-1}$, 51  $\pm$ 28 ~km~s$^{-1}$), 
and ($v_3$, FWHM$_3$ )= (104 $\pm$ 15~km~s$^{-1}$, 221 $ \pm$ 35 ~km~s$^{-1}$).
The integrated flux density of the CO(J=3$\rightarrow$2) emission line is (4.8~$\pm$~1.3~Jy~km~s$^{-1}$)/$\mu_{32}$,
where $\mu_{32}$ is the lensing magnification of the CO(J=3$\rightarrow$2) emitting region estimated in section 6.
We also independently estimated the flux density of the CO(J=3$\rightarrow$2) emission line using the \verb+uvmodelfit+ command that fits a single component source model to the uv data. Using \verb+uvmodelfit+  with the argument \verb+comptype+ set for a Gaussian component model we find an average flux density of I = (8.4 $\pm$ 0.2 mJy)/$\mu_{\rm line}$ within a spectral window of width 544 km~s$^{-1}$.  The best fit-parameters of the properties of the CO(J=3$\rightarrow$2) lines obtained with the extended and compact configurations
are consistent within the error bars. 

The CO(J=2$\rightarrow$1) spectrum was initially fit with a model consisting of a single Gaussian.  This fit is not acceptable in a statistical sense with
 ${\chi^2}/\nu = 39.5/15$, where $\nu$ are the degrees of freedom. 
We next use a model consisting of three Gaussians. The fit with three Gaussians
results in an acceptable fit with  ${\chi^2}/\nu = 11.1/9$.
The best-fit parameters for the velocity centroids and FWHM of the CO(J=2$\rightarrow$1) line were found to be 
($v_1$, FWHM$_1$) = ($-$127 $\pm$ 22~km~s$^{-1}$, 230  $\pm$ 54~km~s$^{-1}$),  
($v_2$, FWHM$_2$) = (10 $\pm$ 7~km~s$^{-1}$, 45  $\pm$ 16~km~s$^{-1}$), 
and ($v_3$, FWHM$_3$ )= (130 $\pm$ 19~km~s$^{-1}$, 174 $ \pm$ 42~km~s$^{-1}$).
The integrated flux density of the CO(J=2$\rightarrow$1) emission line is (2.4~$\pm$~0.7~Jy~km~s$^{-1}$)/$\mu_{21}$, where $\mu_{21}$ is the lensing magnification of the CO(J=2$\rightarrow$1) emitting region. Using the \verb+uvmodelfit+ task we independently estimate an average flux density of I = (3.4 $\pm$ 0.1 mJy)/$\mu_{21}$ within a spectral window of width 638 km~s$^{-1}$.

\section{Lensing Analysis} \label{sec:lensmodel}
\begin{figure}
\includegraphics[width=\columnwidth]{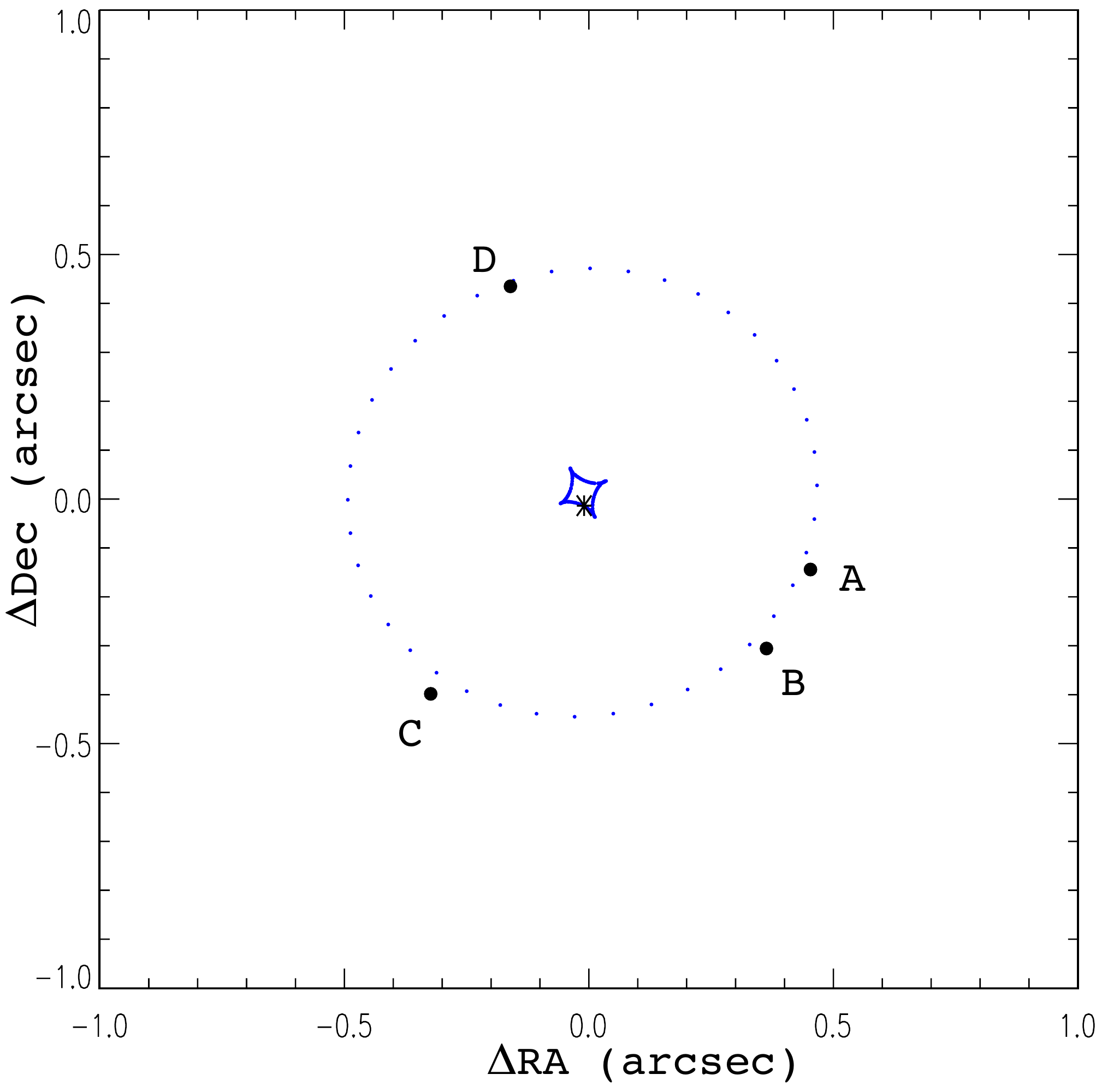}
\caption{Results from lens modeling the HST observation of \hs. The lens model reproduces the observed HST image positions.
This best-fit lens model is used for subsequent lensing analysis of the ALMA observations.
The tangential critical curve is represented by the dotted line, the caustic by the solid line and the UV point source in the source plane with an asterisk.
\label{fig:hst_model}}
\end{figure}

We used the gravitational lens adaptive-mesh fitting code {\sl glafic} \citep{2010ascl.soft10012O} to model
the gravitational lens system \hs. 
{For modeling the source of the UV emission we assumed a point source. This assumption is justified by the estimated size of the UV accretion disk of \hs\ of $r_{\rm s}$ $\sim$ 3 $\times$ 10$^{15}$ cm $\sim$ 0.001 pc, assuming a black hole mass for \hs\ of $M_{\rm BH}$~$\sim$~(4.2 $\pm$ 2.0)~$\times$ 10$^{8}$~$M_{\odot}$ (e.g., \cite{2011ApJ...742...93A})
and the $M_{\rm BH}$ versus $r_{\rm s}$ relation (Figure 9 of \cite{2018ApJ...869..106M}). The HST image of \hs\ does not show an Einstein ring in the UV band indicating that most of the observed UV emission most likely originates from the accretion disk.}
The parameters of the lens are constrained using the optical positions of the images obtained from {\sl HST} observations of \hs. 
The optical positions were taken from the CfA-Arizona Space Telescope LEns Survey (CASTLES) of gravitational lenses website 
\verb+http://cfa-www.harvard.edu/glensdata/+.
The lens is modeled with a singular isothermal ellipsoid (SIE) plus an external shear from the nearby galaxy group.
The ellipsoid's orientation and ellipticity were left as free parameters. The magnification caustics with overlaid source and image positions
are shown in Figure \ref{fig:hst_model}.

\begin{figure}
\includegraphics[width=\columnwidth]{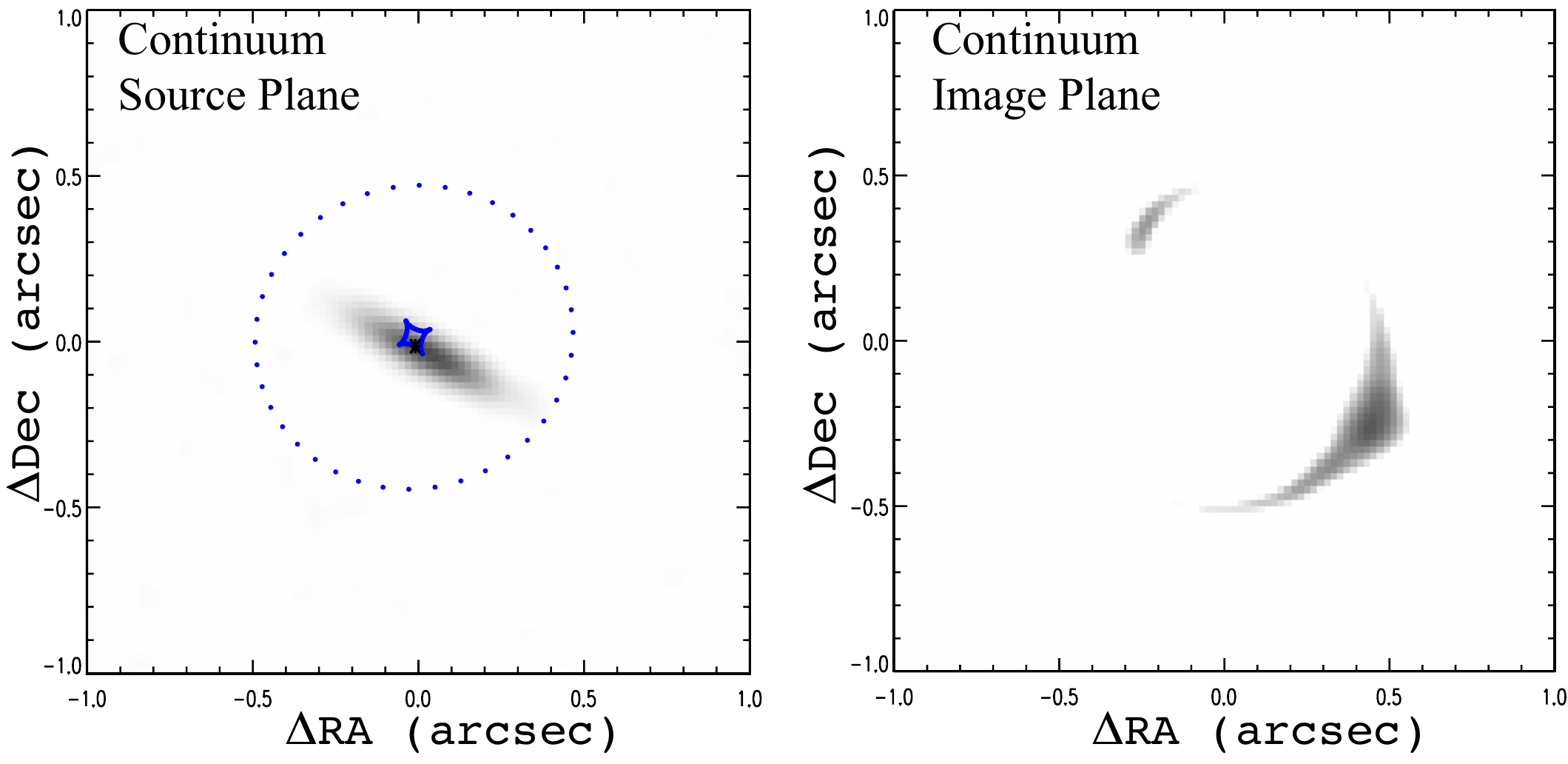}
\caption{Results from lens modeling the $\sim$ 2~mm continuum emission of \hs. 
(left) The $\sim$ 2~mm Continuum emission in the source plane,
(right) The $\sim$ 2~mm Continuum emission in the image plane. 
\label{fig:alma_cont}}
\end{figure}

\begin{figure}
\includegraphics[width=\columnwidth]{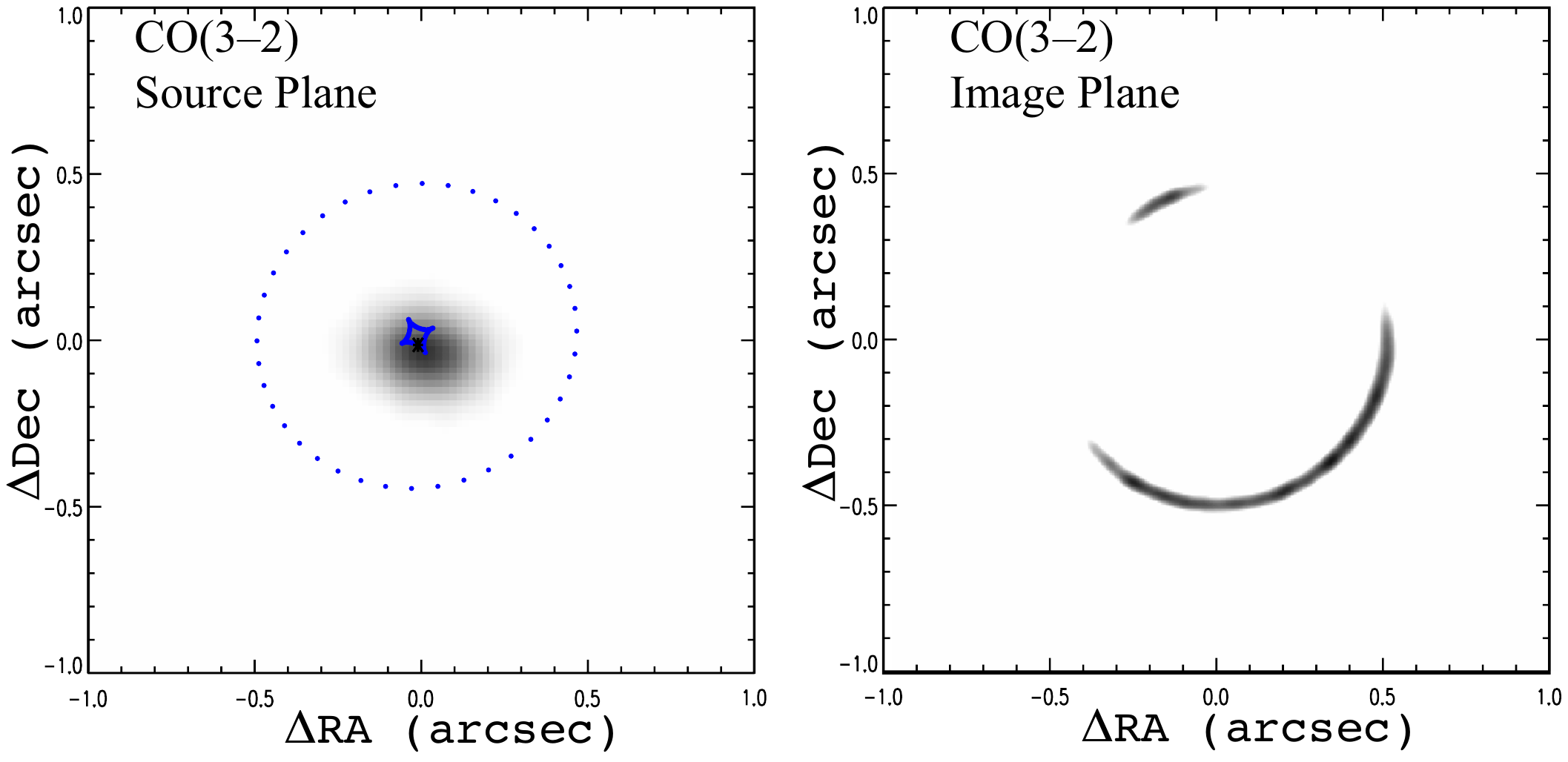}
\caption{Results from lens modeling the CO(J=3$\rightarrow$2) line emission of \hs. 
(left) The CO(J=3$\rightarrow$2) line emission in the source plane,
(right) The CO(J=3$\rightarrow$2) line emission in the image plane. 
\label{fig:alma_line}}
\end{figure}

\begin{figure}
\includegraphics[width=\columnwidth]{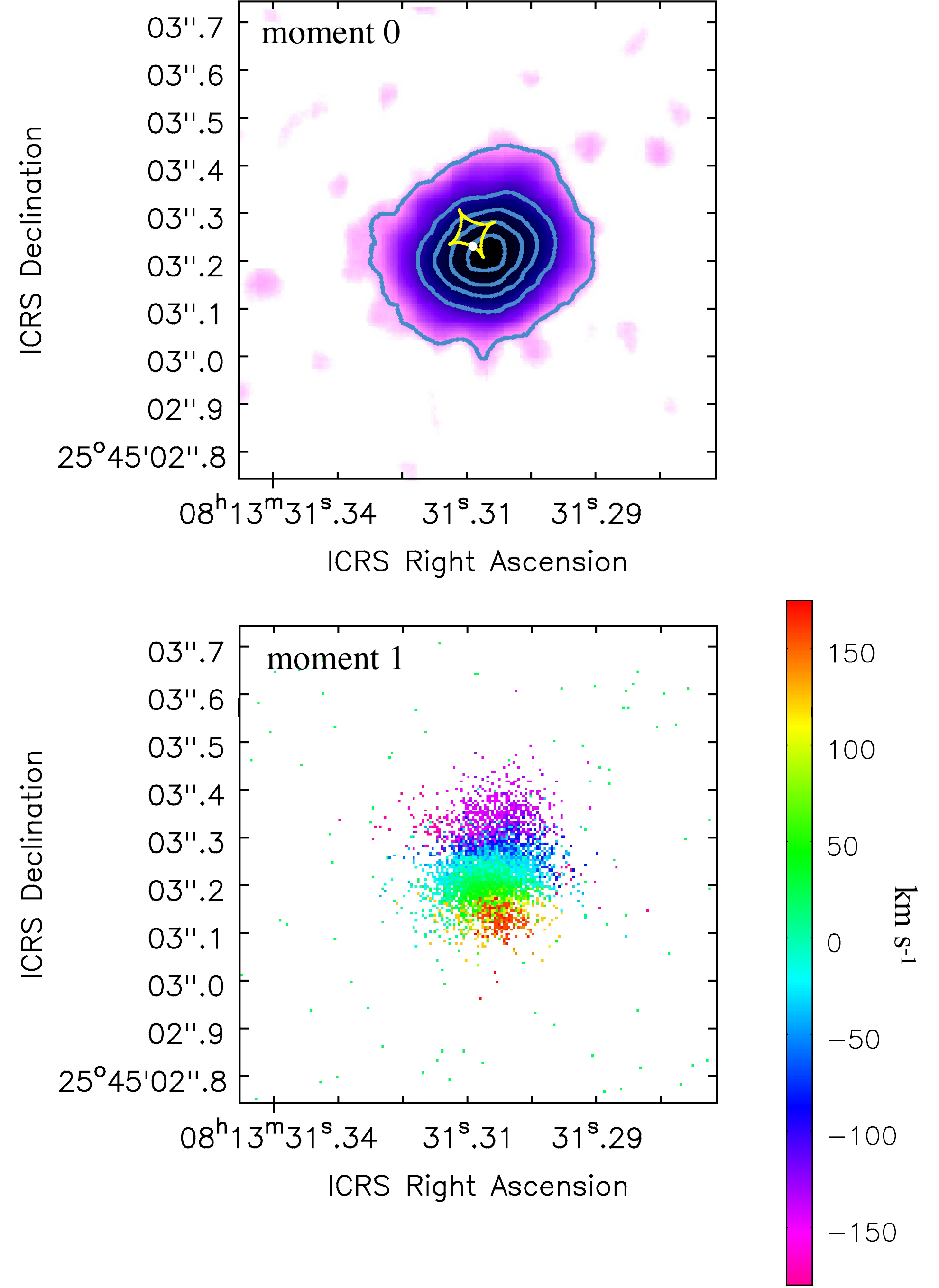}
\caption{
(Top) The  CO(J=3$\rightarrow$2) moment 0 line emission in the source plane, derived from a lens inversion to each frequency channel across the CO(J=3$\rightarrow$2) line (see text for details). 
Contours correspond to 0.9, 0.7, 0.5, 0.3 and 0.1 of the peak. The UV point source in the source plane is represented with a solid white circle. 
(bottom) The moment 1 velocity map of the CO(J=3$\rightarrow$2) line emission of \hs\ calculated in the source plane.
A clear rotation is seen, with the north(south) portions of the CO(J=3$\rightarrow$2) emission moving towards(away from) us.
The rotation is consistent with the spectrum of \hs\ shown in Figure \ref{fig:alma_co32_linespec}.
\label{fig:rgb_source}}
\end{figure}

\begin{figure*}
\includegraphics[width=17.5cm]{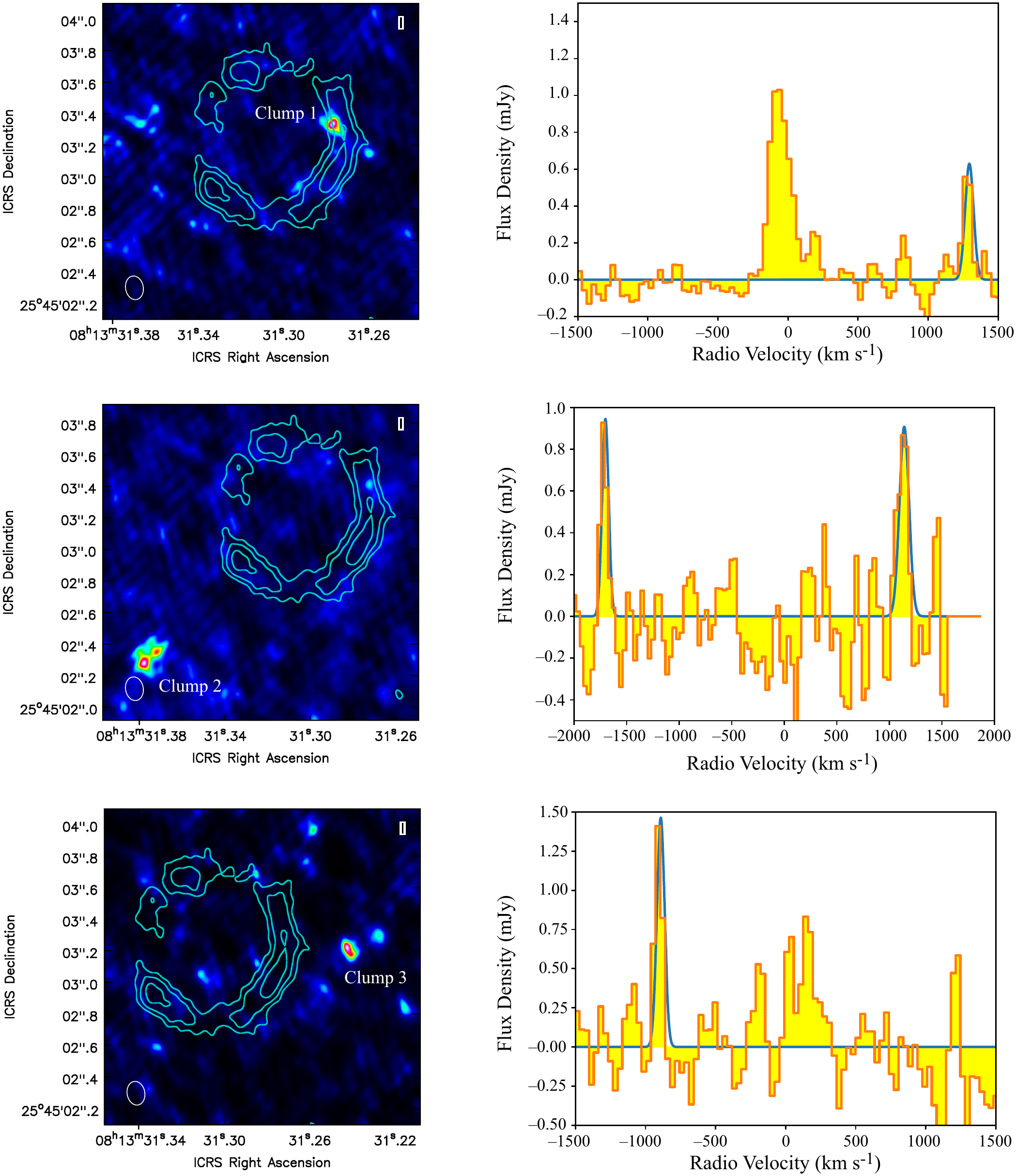}
\caption{Possible blueshifted and redshifted CO(J=3$\rightarrow$2) emission lines associated with clumps. 
(left) Images of \hs\ spectrally integrated over the frequencies within the blueshifted or redshifted emission lines associated with clumps overlaid on the moment 0 CO(J=3$\rightarrow$2) contour map. (right) The ALMA mm spectra extracted from circular regions of radii $\sim$ 0\sarc1 centered on the clumps. 
The best-fit Gaussian-line models are overplotted on the spectra.
\label{fig:alma_clumps}}
\end{figure*}

\begin{figure}
\includegraphics[width=8cm]{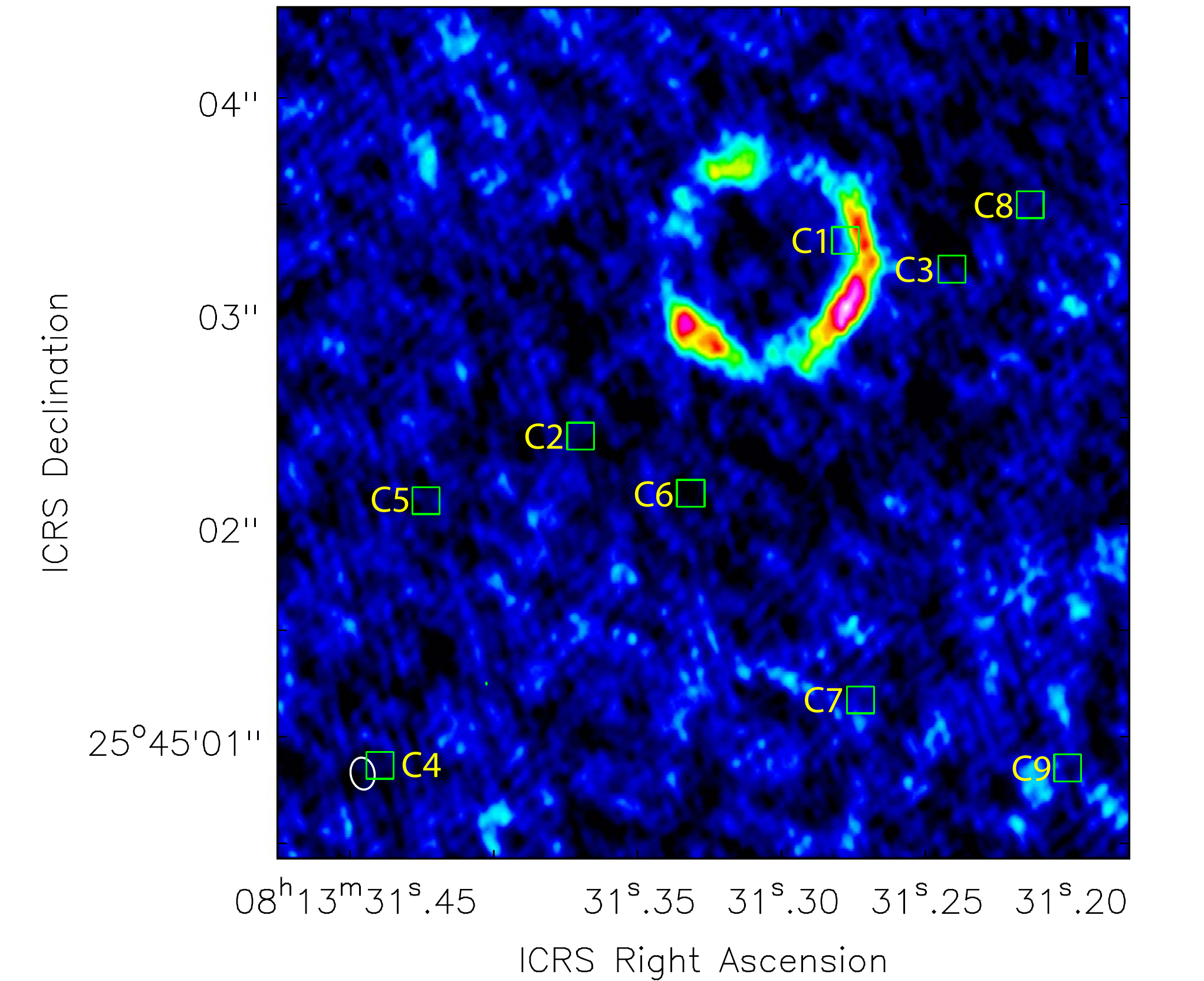}
\caption{{The CO(J=3$\rightarrow$2) moment 0 line emission of \hs\ with the locations of  all the clumps centered on the labelled boxes.
The ALMA pointing was centered on the center of this image and the  Maximum Recoverable Scale  (MRS) in the ALMA extended configuration is MRS~=~1\sarc7. }
\label{fig:all_alma_clumps}}
\end{figure}

In subsequent lens modeling of the ALMA observations we freeze the lens model parameters to the ones constrained from the HST observation. As shown in Figure \ref{fig:hst_alma}, the ALMA band 4 (1.96 mm $-$ 2.19 mm) continuum emission shows a very different morphology than the  $HST$ ACS F555W image of \hs. In particular, the $\sim$ 2~mm continuum emission near images A and B is extended and not aligned with the optical emission and the $\sim$ 2~mm flux of image C is significantly lower than that of image D which is opposite to what is detected in the optical.

We first attempted to model the ALMA  $\sim$ 2~mm continuum emission with a point source, with its position set free in the source plane, assuming our best-fit {\it HST} lens model. The fit with a point source model did not converge indicating that the observed $\sim$ 2~mm continuum emission likely does not originate from a point-like source.  We next model the ALMA $\sim$ 2~mm continuum emission with an extended source in the form of an elliptical gaussian of ellipticity $e$, position angle $\theta_{\rm e}$ and widths along the major and minor axes of  $\sigma_{x}$ and $\sigma_{y}$. As input to the modeling we provide the ALMA high spatial resolution continuum moment 0 image via the {\sl glafic} command \verb+readobs_extend+ and optimize the lens modeling by including a circular mask region with a radius of 0\sarc8  centered on the ALMA emission.  Regions of the ALMA images within this masked region were used for optimization and those outside ignored. 
The {\sl glafic} command \verb+readpsf+ is used to read in the point spread function (PSF) produced by the CASA task \verb+tclean+. The extended source images produced by {\sl glafic} are convolved with the PSF.

In Figure \ref{fig:alma_cont} we show the ALMA $\sim$ 2~mm continuum emission in the source plane and the corresponding lensed image of the ALMA $\sim$ 2~mm continuum emission in the image plane.
The best-fit parameters of the extended continuum source model are presented in Table \ref{tab:lensmodel}. For our assumed cosmology, the FWHM values of the best-fit elliptical gaussian source model to the 2~mm continuum emission along the major and minor axes are  $\sim$1.6~kpc and $\sim$365~pc, respectively. Assuming the $\sim$~2~mm continuum emission originates from an inclined circular disk, we use our source model to derive the disk inclination angle from the minor and major axis ratio (R) (i.e., $i$ = cos$^{-1}$($R$)). The estimated inclination angle of the continuum disk is $i_{\rm cont}$ $\sim$ 77$^{\circ}$. 
{A second possibility is that the mm-continuum emission contains a jet component leading to the highly elongated morphology in the source plane shown in
Figure  \ref{fig:alma_cont}.  VLBI  observations of \hs\ at 1.75~GHz by \cite{2019MNRAS.485.3009H} show a radio jet pointed in a similar direction in \hs. 
Specifically, based on Figure 6 of \cite{2019MNRAS.485.3009H} the position angle of the 1.75~GHz VLBI  jet is $\theta$ = 61$^{\circ}$$^{+10}_{-8}$, which is consistent with the position angle of  $\theta_{e}$  $\sim$ 65 $^{\circ}$ (see Table 3) of the lens inverted $\sim$143~GHz continuum emission.
We investigate the jet hypothesis further by determining the spectral index of the mm-continuum emission of \hs.
Typically, non-thermal synchrotron emission from jets dominates their SED at frequencies $\lesssim$ 10~GHz (\cite{1988AJ.....96...81D}, \cite{1982A&A...116..164G}, \cite{2019A&A...630A..83Z})
and thermal emission from dust in the disk begins to dominate at mm-wavelengths.
The non-thermal synchrotron spectra of jets are often fit with power-law models of the form ${ S }_{ \nu  }\propto { \nu  }^{ \alpha  }$, where $\alpha$ is the spectral index.
The best-fit spectral index of the mm-continuum emission obtained from the ALMA observations of \hs\ is $\alpha$~=~2.8~$\pm$~0.3 which is consistent with
thermal emission and not pure synchrotron emission. We note, however, that strongly inverted spectra with $\alpha > 2$ have been detected in the core of the jet in M87 \citep{2019Galax...7...86Z} and are thought to arize from synchrotron self absorption and free-free absorption from an outflowing wind.
In conclusion, the inverted spectral index of the mm-continuum emission does not support pure synchrotron emission as the origin of the elongated mm-continuum emission detected in the ALMA observations of \hs. As we discuss in section 8, the SED of \hs\ published in \cite{2018MNRAS.476.5075S} is found to have a hot dust component suggesting that an AGN component may be contributing to the sub-mm and mm emission.}

We next apply our lens model to the high resolution CO(J=3$\rightarrow$2)  line image of \hs.
As input to {\sl glafic} we provide the ALMA high spatial resolution CO(J=3$\rightarrow$2) moment 0 image and optimize the lens modeling by including a circular mask region with a radius of 0\sarc8  centered on the ALMA emission. In Figure \ref{fig:alma_line} we show the ALMA CO(J=3$\rightarrow$2) emission in the source plane and the corresponding lensed image of the CO(J=3$\rightarrow$2) emission in the image plane.
We note, however, that the morphology of the CO(J=3$\rightarrow$2) emission in the image and source plane differ in each frequency channel across the CO(J=3$\rightarrow$2) line.
To account for the frequency dependence of the morphology of the CO(J=3$\rightarrow$2) emission and to illustrate the rotation of the CO(J=3$\rightarrow$2) region in the source plane we perform a lens inversion to each frequency channel across the CO(J=3$\rightarrow$2) line 
of \hs\ shown in Figure  \ref{fig:alma_co32_linespec}. Specifically, we first produce images of the CO(J=3$\rightarrow$2) line emission at the thirteen frequency channels between $-$180 km~s$^{-1}$ and 230~km~s$^{-1}$. 
We next use \verb+glafic+ to model each one of the thirteen images independently, assuming lens model parameters obtained from modeling the HST observation of \hs\ as described in $\S$ 6. 
The extended source of the CO(J=3$\rightarrow$2) line emission in each frequency channel is modeled with an elliptical gaussian of ellipticity $e$, position angle $\theta_{\rm e}$ and widths along the major and minor axes of  $\sigma_{x}$ and $\sigma_{y}$, respectively.  We create a spectral cube containing the reconstructed sources at each frequency channel. In Figure  \ref{fig:rgb_source} we show the derived moment 0 spectrally integrated map and the moment 1 velocity map of the CO(J=3$\rightarrow$2) line emission in the source plane. A clear rotation is seen in the moment 1 map, with the north(south) portions of the CO(J=3$\rightarrow$2) emission moving towards(away from) us. The FWHM values of the best-fit elliptical gaussian source model to the CO(J=3$\rightarrow$2) emission along the major and minor axes are $\sim$950~pc and $\sim$690~pc, respectively. Assuming the CO(J=3$\rightarrow$2) line emission originates from an inclined circular disk, our source model implies an inclination angle of the  CO(J=3$\rightarrow$2) disk of $i_{\rm CO(3-2)}$ $\sim$ 43$^{\circ}$

We invoke the {\sl writelens} command in {\sl glafic} to obtain various lensing properties including the magnification map $\mu(i,j)$ of the lens. We use this magnification map to estimate the total magnification of the mm continuum and line emission of \hs\ detected in the ALMA observations.  Specifically, for an observed intensity distribution of $I(i,j)$, the total magnification of the extended source will be

\begin{equation}
{ \mu  }_{ tot }=\frac { \sum { I\left( i,j \right)  }  }{ \sum { \frac { I\left( i,j \right)  }{ \mu \left( i,j \right)  }  }  } 
\end{equation}

We find the total magnifications of the mm-continuum and the CO(J=3$\rightarrow$2) emission to be $\mu_{\rm cont}$ $\sim$ 7 $\pm$ 3 and $\mu_{\rm CO(3-2)}$ $\sim$ 10 $\pm$ 2, respectively.
The calculations of the total magnifications were based on the high spatial resolution observations.
The uncertainty of the magnifications were estimated by considering emission lying within the 2$\sigma$ and 2.5$\sigma$ confidence levels.

{Our lensing analysis indicates that the centroids of the extended mm (continuum and line emission) and point-like UV source emission regions are separated by about 0\sarc35 $\sim$ 300 pc (see Figures 9, 10 and 11). One possible explanation for this offset is that a portion of the extended mm emission is obscured along our line of sight.
We caution, however, that higher S/N ALMA images would be required to confirm this offset.}

\section{Possible detection of highly blueshifted and redshifted clumps of CO(J=3$\rightarrow$2) emission} \label{sec:outflow_inflow}
We have identified ten blueshifted/redshifted emission lines that are associated with nine clumps of mm emission. Given the relatively high velocities observed ($\simgt$1000 km~s$^{-1}$), it is unlikely that the redshifted emission is produced by the doppler shift of infalling gas from the near side of the CO gas. 
A more likely scenario is that the blueshifted (redshifted) emission is produced by the  doppler shift of outflowing material from the near(far) side of the CO gas. 
In the left panels of Figure \ref{fig:alma_clumps}  we show the images of \hs\ spectrally integrated  over the frequencies within the blueshifted or redshifted emission lines associated with clumps overlaid on the moment 0 CO(J=3$\rightarrow$2) contour map. On the right panels of Figure \ref{fig:alma_clumps} we show the mm spectra extracted from circular regions of radii $\sim$ 0\sarc1 centered on the clumps.
The properties and significance of the blueshifted and redshifted emission lines were determined by fitting them with Gaussians and calculating the
integrated rms of the background under each line.
{The  significance of the lines detected lie in the range of 3.0$-$4.7$\sigma$ and the velocities lie between $-$1702~km~s$^{-1}$ and $+$1304~km~s$^{-1}$.
In Table \ref{tab:clumps} we list the positions, velocities with respect to systemic, FWHM, integrated flux densities with $1\sigma$ errors, detection significance, lensing magnifications, distances from the center of the galaxy and total masses of the clumps.  
{In Figure \ref{fig:all_alma_clumps} we show the spatial distribution of all the clumps overplotted on the CO(J=3$\rightarrow$2) moment 0 line emission of \hs.}
We note that clump 1 is also clearly visible in the pv diagram extracted along the A$+$B  images and shown in Figure \ref{fig:alma_clumps}. }

\section{Discussion and Conclusions} \label{sec:conclusions}

We have presented results from the spectral and spatial analysis of ALMA observations of the $z$ = 1.51 lensed quasar \hs.
Several important properties of the molecular gas in the host galaxy can be inferred from these results.
Specifically, we are able to estimate both the total mass of the molecular gas and the energetics of the possible outflows associated with possible high velocity clumps of molecular gas detected in \hs.

\begin{figure}
\includegraphics[width=\columnwidth]{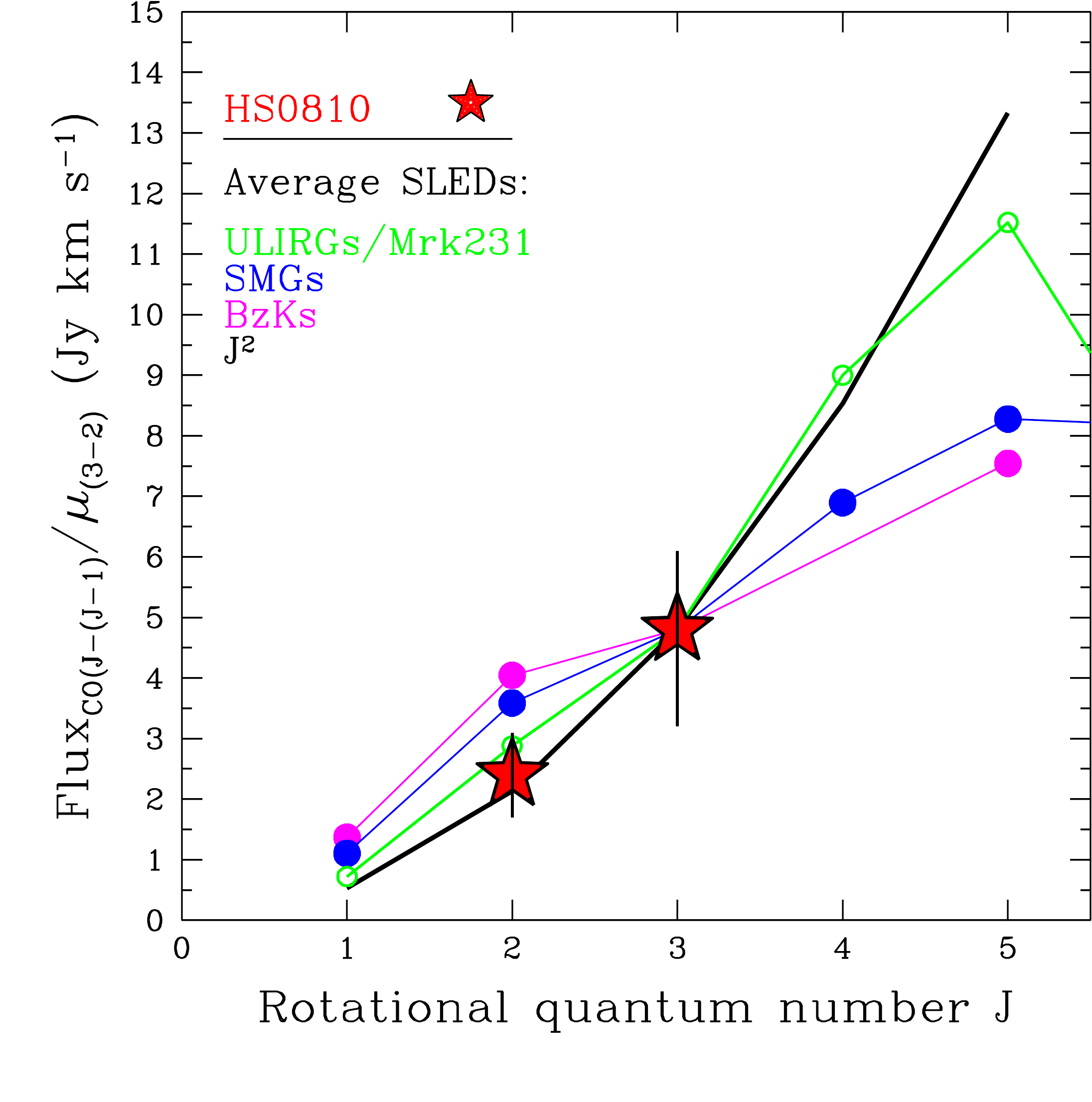}
\caption{CO excitation ladder of \hs\ (stars) compared with average SLED values obtained for various classes of sources (SMGs, ULIRGs, BzKs, as labeled; taken from Papadopoulos et al. (2012), Bothwell et al. (2013), Daddi et al. (2015)), for Mrk 231 (from Fixsen et al. (1999) and van
der Werf et al. (2010)). The black curve show the J$^2$ trend (e.g. LTE assumption).  All CO SLEDs are normalized to the CO(J=3$\rightarrow$2) flux observed in \hs.
\label{fig:sled}}
\end{figure}

In section 5 we show that the integrated flux density of the CO(J=2$\rightarrow$1) and CO(J=3$\rightarrow$2) emission of \hs\ are 2.4 $\pm$ 0.7 Jy~km~s$^{-1}$/$\mu_{21}$  and 4.8 $\pm$ 1.3 Jy~km~s$^{-1}$/$\mu_{32}$, respectively.
Figure \ref{fig:sled} shows the CO Spectral Line Energy Distribution (SLED) of \hs\ based on these two values and compared to several CO SLEDs of different classes of high-$z$ and ultraluminous galaxies. The ratio of the CO(J=3$\rightarrow$2)  to CO(J=2$\rightarrow$1)  emissions measured from our ALMA data shows that the CO SLED of \hs\ is consistent with a J$^{2}$ dependence applicable to bright QSOs from \cite{2013ARA&A..51..105C}. 
{Our lensing analysis in section 6 describes our method for estimating the magnification factors $\mu_{32}$ = 10 $\pm$ 2 and $\mu_{\rm cont}$ = 7 $\pm$ 3.
Assuming the geometries of the CO(J=2$\rightarrow$1)/$\mu_{21}$ and CO(J=3$\rightarrow$2)/$\mu_{32}$ emission regions are similar we adopt $\mu_{32}$ = $\mu_{21}$.}

The CO line luminosity can be expressed as,

\begin{equation}
{ L }_{\rm CO }=2453{ S }_{\rm CO }\Delta v{ D }_{\rm L }^{ 2 }{ \left( 1+z \right)  }^{ -1 }
\end{equation}

\noindent
{where, ${ S }_{\rm CO }\Delta {v}$ is the integrated flux density of the CO(J=1$\rightarrow$0) line. Assuming  magnifications of $\mu_{32}$ = $\mu_{21}$ = 10 $\pm$ 2, and 
${ S }_{\rm CO(3-2) }\Delta {v}$/${ S }_{\rm CO(1-0) }\Delta {v}$ = 9 (see Figure \ref{fig:sled})
we find ${ L }_{\rm CO }$~=~(6.6~$\pm$~1.9)/$\mu_{32}$~$\times$~10$^{10}$ K~km~s$^{-1}$~pc$^{2}$.}

Following \cite{2005ARA&A..43..677S} the total mass of the molecular gas is expressed as,
\begin{equation}
{ M }_{\rm Mol }={\alpha}_{\rm CO } { L }_{\rm CO }
\end{equation}

\noindent
where, ${ M }_{\rm Mol }$ is in units of $M_{\odot}$, ${ L }_{\rm CO }$ is in units of K~km~s$^{-1}$~pc$^{2}$, and ${\alpha}_{\rm CO }$ is the CO$-$to$-$H$_{2}$ conversion factor.
{Assuming a ULIRG-like CO$-$to$-$H$_{2}$ conversion factor of $\alpha_{\rm CO}$~=~0.8 $M_{\odot}$~(K~km~s$^{-1}$~pc$^{2}$)$^{-1}$ we find that the total gas mass of the molecular gas in \hs\ to be ${ M }_{\rm Mol }$  = (5.2 $\pm$ 1.5)/$\mu_{32}$ $\times$ 10$^{10}$~$M_{\odot}$.}

{
\cite{2018MNRAS.476.5075S} note that the high effective dust temperature of $\sim$ 89 K of \hs\ may be the result of a significant contribution from the AGN heating the dust in addition to heating provided by star formation. By modeling the FIR dust emission with a 2 temperature model Stacey et al. derive a FIR luminosity of 3.7$^{+1.7}_{-2.3}$ $\times$ 10$^{12}$ $L_{\odot}/\mu_{\rm FIR}$ for the cold component, that likely originates from star formation.
Assuming the FIR luminosity of the cold component is produced by star formation, we derive a star formation rate (SFR) for \hs\ of SFR~=~640$^{+400}_{-300}/\mu_{\rm FIR}$~$M_{\odot}$~yr$^{-1}$ \citep{1998ApJ...498..541K}.
We estimate the molecular gas depletion time, defined as,

\begin{equation}
{ \tau  }_{\rm dep }^{\rm mol }=\frac { { \Sigma  }_{\rm mol } }{ { \Sigma  }_{\rm SFR } }
\end{equation}

\noindent
to be ${ \tau  }_{\rm dep }^{\rm mol }$ $\sim$ 8.2 $\times$ 10$^{7}$~$(\mu_{\rm FIR}/\mu_{32})$~yr, where
${ \Sigma  }_{\rm mol }$ and ${ \Sigma  }_{\rm SFR }$ are the surface densities of molecular gas and star formation, respectively.}

Assuming the CO(J=3$\rightarrow$2) line emission originates from an inclined rotating circular disk we estimate the dynamical mass of this disk from the relation $M_{\rm dyn}$~=~1.16~$\times$~10$^{5}$$v_{\rm cir}^{2}D$~$M_{\odot}$, where $D$ is the disk diameter in kpc and $v_{\rm cir}$ is the maximum circular velocity of the CO(J=3$\rightarrow$2) disk in km~s$^{-1}$ \citep{2013ApJ...773...44W}. $v_{\rm cir}$ is approximated as  $v_{\rm cir}$ = 0.75FWHM$_{CO(3-2)}$/sin$i$, where $i$ is the inclination angle of the disk.  We estimate the dynamical mass within the CO(J=3$\rightarrow$2) disk to be $M_{\rm dyn}$~=~(3.2~$\pm$~0.4)~$\times$~10$^{10}$~$M_{\odot}$.

In Table \ref{tab:clumps} we also provide estimates of the mass outflow rates associated with the redshifted and blueshifted clumps.
The masses in the clumps range from 0.3 $\times$ 10$^{8}$ to 8 $\times$ 10$^{8}$ $M_{\odot}$ (of the order of few \% of the total gas mass).

 Assuming that the blueshifted and redshifted clumps are associated with outflows, we provide an estimate of the outflow rates averaged over the wind lifetime of the detected clumps based on the equation \citep{2005ApJ...631L..37R}:

\begin{equation}
\dot { { M } } = \frac { { M }_{\rm clump }{ v }_{\rm clump } }{ { r }_{\rm clump } } 
\end{equation}

This equation assumes that the clumps have been moving at the observed velocity of ${v }_{\rm clump}$ for their lifetime.
We note that ${ r }_{ clump}$ in this equation is the distance of a clump from the AGN, whereas, we observe projected distances.   
Our estimates of the total mass of the molecular gas and the mass inflow and outflow rates of \hs\ are used to determine the dynamical time, $t_{\rm dyn}$, in which the gas can be depleted. Specifically,  $t_{\rm dyn}$ is derived from the expression,

\begin{equation}
{ t }_{\rm dyn }\approx \frac { { M }_{\rm Mol } }{  { \dot { M }  }_{\rm Outflow }   } 
\end{equation}

We find a dynamical depletion time of ${ t }_{\rm dyn}$ = (1.3~$\pm$~0.4)/$\mu_{32}$~$\times$~10$^{8}$ years.

The rate of change of momentum of the molecular outflow is $\dot { { p } }_{\rm mo}=\dot {{M}} v_{\rm clump}$ and the momentum boost is 
$\dot { { p } }_{\rm mo}/(L_{Bol}/c)$. The bolometric luminosity, $\log(L_{\rm Bol}/\rm erg~s^{-1})$ = 45.54 (corrected for magnification of $\mu = 103$), is provided from the monochromatic luminosity of \hs\ at 1450\AA, $\log({\lambda}L_{\lambda}/\rm erg~s^{-1})$ = 45.04, based on the empirical equations of \cite{2012MNRAS.422..478R}. The UV monochromatic luminosity density of \hs\ at 1450{\AA} was obtained from analyzing its available SDSS spectrum.

\begin{figure}
\includegraphics[width=\columnwidth]{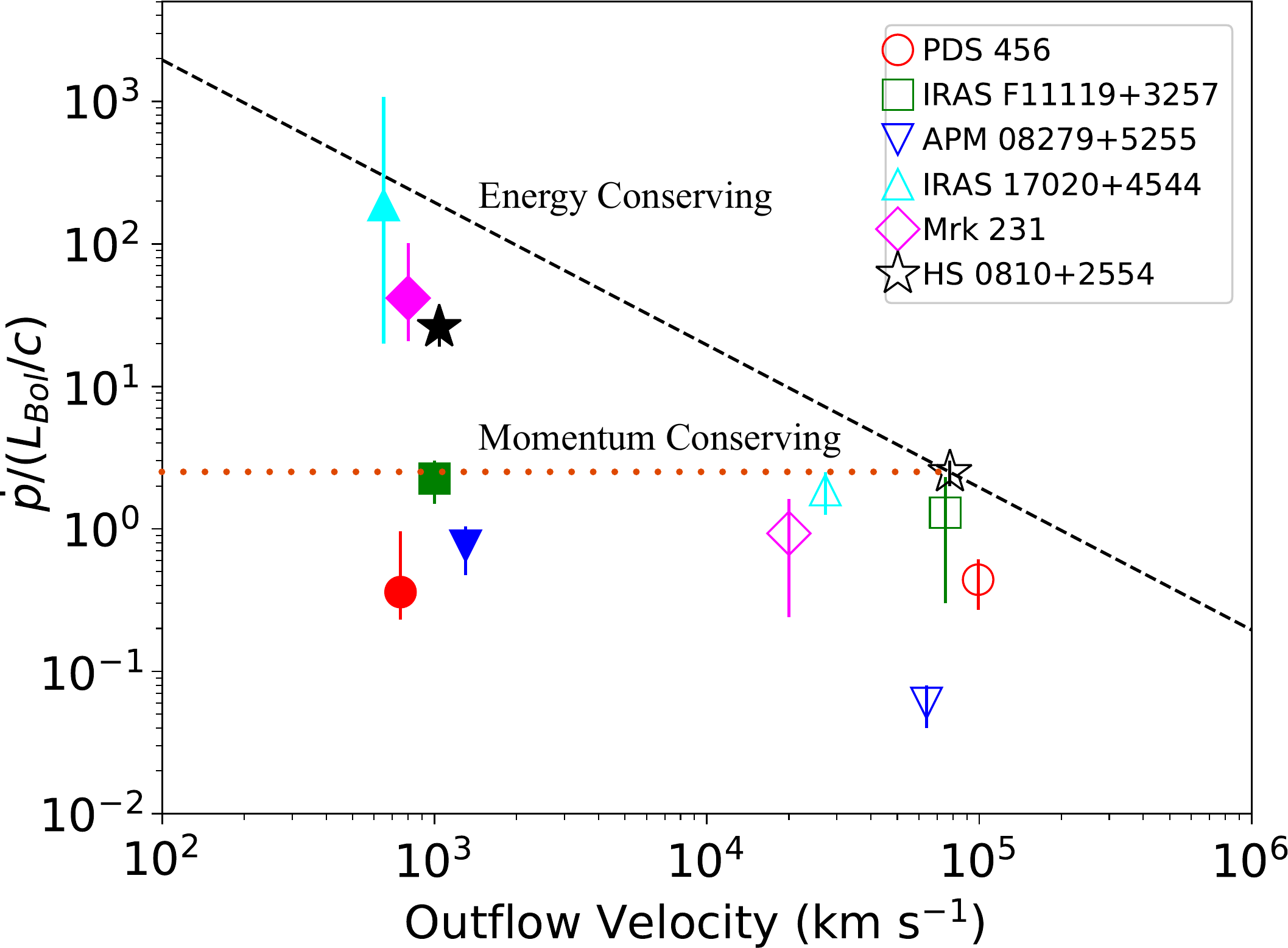}
\caption{The momentum boost $\dot { { p } }/(L_{Bol}/c)$ as a function of outflow velocity. 
The filled and unfilled symbols correspond to the molecular and ultrafast outflows, respectively. The dashed and dotted lines represent the dependence of the momentum boost with outflow velocity for energy-conserving and momentum-conserving outflows, respectively for \hs. 
The data for AGN other than \hs\ have been obtained from the literature (eg., Bischetti et al. 2019, Tombesi et al. 2015, Veileux et al 2017, Feruglio et al. 2017, Feruglio et al. 2015).
For consistency we have adjusted published results assuming the same conversion factor of $\alpha_{\rm CO}$~=~0.8 $M_{\odot}$~(K~km~s$^{-1}$~pc$^{2}$)$^{-1}$ for estimating the total molecular gas mass.
\label{fig:momentum_boost}}
\end{figure}

In Figure \ref{fig:momentum_boost} we show the momentum boost $\dot { { p } }/(L_{Bol}/c)$ plotted against the outflow velocity for  \hs\  based on this work. The ($\dot { { p } }_{\rm xo}/(L_{Bol}/c)$, $v_{\rm xo}$) values for the two components of the ultrafast outflow of \hs\ obtained from the observation on 2014 October 4 are (1.5 $\pm$ 0.4, 0.12 $\pm$ 0.04 $c$) and (1.0 $\pm$ 0.2, 0.42 $\pm$ 0.04 $c$). 
The combined momentum boost and the average outflow velocity of the ultrafast outflow of \hs\  are shown in Figure \ref{fig:momentum_boost}.
Based on our analysis the ($\dot { { p } }_{\rm mo}/(L_{Bol}/c)$, $v_{\rm mo}$) values of the molecular outflow {are (26~$\pm$~6, 1040~km~s$^{-1}$)}.
The momentum boost of the molecular outflow of \hs\ is calculated by combining the nine outflowing clumps listed in Table \ref{tab:clumps} and the outflow velocity represents the average of the clump velocities. The momentum boost of the outer molecular wind of \hs\ is slightly below the value expected for an energy-conserving outflow, given the observed ultrafast outflow (UFO) momentum boost.

In Figure \ref{fig:momentum_boost} we also show the momentum boosts  and velocities of the ultrafast and molecular outflows of several other AGN based on published results.
Specifically, estimates for PDS~456 are from \cite{2019ApJ...873...29B} and \cite{2019A&A...628A.118B}, estimates for IRAS~F11119+3257 are based on \cite{2015Natur.519..436T} and \cite{2017ApJ...843...18V},
estimates for APM~08279+5255 are based on \cite{2017A&A...608A..30F}, estimates for  IRAS~17020$+$4544 are from \cite{2018ApJ...867L..11L}, estimates for Mrk~231 are from \cite{2015A&A...583A..99F},
and estimates for \hs\ are based on \cite{2016ApJ...824...53C} and this study.
For consistency we have adjusted published results assuming the same conversion factor of $\alpha_{\rm CO}$~=~0.8 $M_{\odot}$~(K~km~s$^{-1}$~pc$^{2}$)$^{-1}$ for estimating the total molecular gas mass. The two AGN of this small sample that have momentum boosts consistent with a momentum-conserving outflow are PDS~456 and  IRAS~F11119+3257, whereas the other objects including \hs\ are  closer to having energy-conserving outflows. \cite{2019ApJ...887...69S} find a similar result in the sense that the 
efficiency factors coupling the small scale relativistic winds and the large scale molecular outflows of their sample of objects varies significantly.  

{
\cite{2019ApJ...871..156M} reported several trends between the energy-transfer rate $C$ defined as,

\begin{equation}
C=\frac { \frac { 1 }{ 2 } { \dot { M }  }_{\rm  mol }{ v }_{\rm mol }^{ 2 } }{ \frac { 1 }{ 2 } { { \dot { M }  }_{\rm  UFO }v }_{\rm  UFO }^{ 2 } } 
\end{equation}

\noindent
and AGN properties such as black hole mass $M_{\rm BH}$ and $L_{\rm Bol}/L_{\rm Edd}$.
We calculate the energy-transfer rate of \hs\ to be $C$~=~0.2~$\pm$~0.1 with 
$L_{\rm Bol}/L_{\rm Edd}$ = 0.07 $\pm$ 0.03. We find that \hs\ is only marginally consistent, within the error bars, with the $C$ vs. $M_{\rm BH}$  and $C$ vs. $L_{\rm Bol}/L_{\rm Edd}$ trends reported in \cite{2019ApJ...871..156M}.
We estimated the strength of the proposed $M_{\rm BH}$ vs. C and $L_{\rm Bol}/L_{\rm Edd}$ vs. C trends before and after adding the data from \hs.
Before including \hs\ to the \cite{2019ApJ...871..156M} sample we find that the Kendall's rank correlation coefficient, $\tau$ and its significance are $\tau$ = $-$0.33 significant at $>$ 71\% confidence for the $C$ vs. $M_{\rm BH}$  data and $\tau$ = 0.62 significant at $>$ 95\% confidence for the $L_{\rm Bol}/L_{\rm Edd}$ vs. $C$ data.
Including \hs\ to the Mizumoto et al. 2019 sample we find $\tau$ = $-$0.21 significant at $>$ 54\% confidence for the $C$ vs. $M_{\rm BH}$ data
and $\tau$ = 0.57 significant at $>$ 95\% confidence for the $C$ vs. $L_{\rm Bol}/L_{\rm Edd}$  data.
In conclusion, we find no significant trend of the energy-transfer rate vs. $M_{\rm BH}$, whereas we confirm that the trend of the energy-transfer rate vs. $L_{\rm Bol}/L_{\rm Edd}$
is significant at the $>$ 95\% confidence level.}

There are significant limitations in using single epoch momentum boost estimates of molecular and ultrafast outflows to infer whether a flow is energy or momentum conserving. We are currently observing the energetics of the macro-scale molecular outflows that were possibly driven by micro-scale ultrafast outflows produced a considerable time earlier. According to several theoretical models (e.g., \cite{2012MNRAS.425..605F}, \cite{2012ApJ...745L..34Z}), ultrafast outflows are thought to interact with the ISM and then slow down to a speed of $\sim$ 1000 km~s$^{-1}$ and travel at this speed through the galaxy.
The time for these slow molecular outflows to travel a distance of $\sim$ 1$-$10~kpc (observed distances of molecular outflows) is 10$^{6}$ to 10$^{7}$ years.  It is therefore possible that the energetics of the ultrafast outflows have varied significantly over the past 10$^{6}$ to 10$^{7}$ year time frame making the comparison of the energetics of the micro and macro scale outflows
difficult.
Another important issue is that part of the energy of the micro-scale ultrafast outflow may have been channeled to drive ionized gas which has not been included in this analysis.

The main conclusions of our spectral and spatial analyses of the ALMA observations of \hs\ are the following:

\begin{enumerate}

\item{The mm-continuum emission of \hs\ is detected and resolved with ALMA. {The flux density of the continuum of the combined images at a mean frequency of 143~GHz is estimated to be (401 $\pm$  50)/$\mu_{\rm cont}$ $\mu$Jy, where $\mu_{\rm cont}$ = 7 $\pm$ 3 is our estimated lensing magnification of the continuum.
The positions, flux densities and significances of the resolved lensed images are listed in Table 2.}}

\item{The CO(J=2$\rightarrow$1) and  CO(J=3$\rightarrow$2) line emissions of \hs\ are detected with ALMA. The CO(J=3$\rightarrow$2) line emission was observed with the ALMA extended configuration (beam size 0\sarc1 $\times$ 0\sarc06) revealing a spectacular Einstein ring (see Figure \ref{fig:alma_cont_line}). {The integrated flux density of the CO(J=3$\rightarrow$2) emission line is (4.8~$\pm$~1.3~Jy~km~s$^{-1}$)/$\mu_{32}$, where $\mu_{32}$ = 10 $\pm$ 2. The integrated flux density of the CO(J=2$\rightarrow$1) emission line is (2.4~$\pm$~0.7~Jy~km~s$^{-1}$)/$\mu_{21}$, where $\mu_{21}$ = 10 $\pm$ 2. Assuming a ratio of CO(J=3$\rightarrow$2)/CO(J=1$\rightarrow$0) $\sim$ 9 based on CO-SLED model for quasars and for an $\alpha_{\rm CO}$~=~0.8 $M_{\odot}$~(K~km~s$^{-1}$~pc$^{2}$)$^{-1}$ we estimate the molecular gas mass of \hs\ to be 
${ M }_{\rm Mol }$  = (5.2 $\pm$ 1.5)/$\mu_{32}$ $\times$ 10$^{10}$~$M_{\odot}$.}
We estimate the dynamical mass of the CO(J=3$\rightarrow$2) disk to be $M_{\rm dyn}$~=~(3.2~$\pm$~0.4)~$\times$~10$^{10}$~~$M_{\odot}$.

}

\item{We find a significant offset between the positions of the lensed images of \hs\ as observed in the mm-continuum band
and the image positions in the optical (see Figure \ref{fig:hst_alma}). An offset is also detected between the lensed images observed in the mm-continuum band and the Einstein ring that is resolved in the CO(J=3$\rightarrow$2) line (see Figure \ref{fig:alma_cont_line}).
Our lens modeling of \hs\ indicates that the image offset is caused by the different morphologies of the mm-continuum and CO(J=3$\rightarrow$2) line emission regions. Specifically,  based on our lens modeling, we find the FWHM values of the best-fit elliptical gaussian source model to the 2 mm continuum emission along the major and minor axes are $\sim$ 1.6 kpc and $\sim$ 365 pc, respectively. 
{We caution that our constrained values for the size of the extended continuum emission is very uncertain due to the assumed simplistic elliptical source model, the low S/N of the continuum ALMA data and the inherent problems with lens modeling the synthesized ALMA data.
Higher S/N ALMA images of the continuum will be required before a more detailed lensing analysis can be performed to better constrain the source morphology and provide insight into the origin of the $\sim$ 2~mm-continuum source emission.} The FWHM values of the best-fit elliptical gaussian source model to the CO(J=3$\rightarrow$2)  emission along the major and minor axes are $\sim$950~pc and $\sim$ 690~pc, respectively. Assuming the $\sim$ 2~mm continuum and CO(J=3$\rightarrow$2) emission originate from inclined circular disks, our source models imply an inclination angle of the continuum disk of $i_{\rm cont}$ $\sim$ 77$^{\circ}$ and of the CO(J=3$\rightarrow$2) disk of  $i_{\rm CO(3-2)}$ $\sim$ 43$^{\circ}$.
We investigated a second possibility that the continuum emission contains a jet leading to the highly elongated morphology in the source plane shown in Figure  \ref{fig:alma_cont}.  VLBI observations at 1.75~GHz by \cite{2019MNRAS.485.3009H} find a radio jet pointed in a similar direction in \hs. The inverted spectral index of $\alpha$ = 2.8 $\pm$ 0.3 of the mm-continuum emission of \hs,  however, does not support pure synchrotron emission as the origin of the elongated mm-continuum emission.

}

\item{The CO(J=2$\rightarrow$1) and CO(J=3$\rightarrow$2) line spectra of \hs\ obtained with the ALMA compact configuration are broad and appear to contain three dominant peaks. The two outer peaks of the CO(J=2$\rightarrow$1) and  CO(J=3$\rightarrow$2) lines are separated by $\Delta{v_{21}}$ = 257 $\pm$ 29 km~s$^{-1}$ and $\Delta{v_{32}}$ = 228 $\pm$ 27 km~s$^{-1}$, respectively. The CO(J=3$\rightarrow$2) line spectrum of the high spatial resolution dataset also contains three dominant peaks with the two outer peaks separated by $\Delta{v_{32}}$ = 254 $\pm$ 23 km~s$^{-1}$. These velocity shifts imply the presence of a rotating molecular disk. A rotating molecular disk is supported by the detected shift of the CO(J=3$\rightarrow$2) spectrum as a circular spectral extraction region is shifted across the stretched image of the extended line emission (see Figure \ref{fig:alma_co32_doppler}).
Our reconstruction of the CO(J=3$\rightarrow$2) source emission in multiple radio velocities clearly shows the rotation of the molecular gas (see Figure \ref{fig:rgb_source}).  
} 

\item{We report the possible detection of highly redshifted and blueshifted clumps of CO(J=3$\rightarrow$2) emission.
{We assume that the blueshifted (redshifted) emission is produced by the  doppler shift of outflowing material from the near(far) side of the CO gas. 
The significance of the detections of the clumps ranges from 3${\sigma}$ to 4.7${\sigma}$, the velocities range from $-$1702~km~s$^{-1}$  to 1304~km~s$^{-1}$ and the estimated mass outflow rates range from 
$\sim$ 7~M$_{\odot}$~year$^{-1}$ to $\sim$ 75~M$_{\odot}$~year$^{-1}$. }
{We estimate the dynamical depletion time of the molecular gas to be ${ t }_{\rm dyn}$ = (1.3 $\pm$ 0.4)/$\mu_{32}$~$\times$~10$^{8}$ years.}
This time does not include possible depletion of gas due to star formation. 
}

\item{
We would like to point out that while in the local Universe ($z$ $\simlt$ 0.2) the number of objects with simultaneous detection of UFOs and molecular outflows is approaching $\sim$10 (see \cite{2019ApJ...887...69S}, and \cite{2019ApJ...871..156M}, \hs\ is the only other object at $z>1$ in addition to APM~08279+5255 where a simultaneous detection has been made. 
We estimate the momentum boost, $\dot { { p } }_{\rm mo}/(L_{Bol}/c)$, of the possible molecular outflow {of \hs\ to be 26~$\pm$~6.}  
The momentum boost of the molecular wind of \hs\ is slightly below the value predicted for an energy-conserving outflow given the momentum flux observed in the ultrafast outflow. 
We caution that the significance of several of the blueshifted and redshifted clump detections are relatively low and a deeper ALMA observation of \hs\ would be required to confirm these results.}

{
\item{
\cite{2019ApJ...871..156M} reported trends between the energy-transfer rate $C$ and the black hole mass $M_{\rm BH}$ and $L_{\rm Bol}/L_{\rm Edd}$.
With and without including \hs\ to the Mizumoto et al. 2019 sample we find no significant trend of the energy-transfer rate vs. $M_{\rm BH}$, whereas we confirm that the trend of the energy-transfer rate vs. $L_{\rm Bol}/L_{\rm Edd}$ is significant at the $>$~95\% confidence level.

}
}

\end{enumerate}

\section*{Acknowledgements}
We acknowledge financial  support from PRIN MIUR 2017PH3WAT (``Black hole winds and the baryon life cycle of galaxies"). 
GC would like to express his profound appreciation to the gracious faculty and staff of the 
Dipartimento di Fisica e Astronomia dell'Universit\`{a} degli Studi di Bologna and INAF/OAS of Bologna,
for their enduring collaboration and generous hospitality in providing a stimulating environment during his visits to their esteemed institutions.
We greatly appreciate the useful comments made by the referee. 
This paper makes use of the following ALMA data: ADS/JAO.ALMA\#2017.1.01368.S. ALMA is a partnership of ESO (representing its member states), NSF (USA) and NINS (Japan), together with NRC (Canada), MOST and ASIAA (Taiwan), and KASI (Republic of Korea), in cooperation with the Republic of Chile. The Joint ALMA Observatory is operated by ESO, AUI/NRAO and NAOJ.
The National Radio Astronomy Observatory is a facility of the National Science Foundation operated under cooperative agreement by Associated Universities, Inc.






\clearpage
\begin{table}
\caption{Log of ALMA observations of  \hs. \label{tab:obslog}\}} 
\begin{tabular}{llccccccc}
 & & & & & & & & \\ \hline\hline
Date${}^{a}$ & $t_{\rm int}$${}^{b}$ & Beam ${}^{c}$  & SPW${}^{d}$   & width${}^{e}$    & channels${}^{f}$        & Resolution${}^{g}$       & Effective Resolution${}^{h}$ & band  \\
                      & (sec)                        &                         &                          &(GHz)               &                                    & km~s$^{-1}$                 & km~s$^{-1}$                          &      \\
\hline
2017~Nov~30     &  1655   & 0.12\arcsec $\times$ 0.06\arcsec  & 138.0/139.9/150.0/152.0 & 2  & 128 & 34.0 & 34.0  & 4   \\
2018~Jan~28       &  1716   & 1.10\arcsec $\times$ 0.81\arcsec    & ~91.8/~90.2/102.3/104.2 & 1.875 & 1920 & 3.2 &25.6  & 3 \\
2018~Aug~29        &    544         & 1.48\arcsec $\times$ 1.03\arcsec   & 137.8/135.9/149.8/147.9 & 1.875 & 1920 & 2.1 &  8.4&  4 \\
\hline 
\multicolumn{9}{l}{$^a$ Date of start of observation.}\\
\multicolumn{9}{l}{$^b$ Total integration time on source.}\\
\multicolumn{9}{l}{$^c$ The beam size obtained by adopting the Briggs weighting scheme and a robust parameter of R = 2.}\\
\multicolumn{9}{l}{$^d$ The central frequencies of the spectral windows.}\\
\multicolumn{9}{l}{$^e$ The width of each spectral window.}\\
\multicolumn{9}{l}{$^f$ The number of channels in each spectral window.}\\
\multicolumn{9}{l}{$^g$ The velocity resolution at the frequency of the CO line present in the spectral window.}\\
\multicolumn{9}{l}{$^h$ The effective velocity resolution obtained after binning the spectra.}\\
\end{tabular}
\end{table}

\begin{table*}
\caption{ALMA positions and integrated flux densities of the lensed images (A, B, C, and D) of the mm-continuum of \hs. For comparison we also list the relative HST positions of the lensed optical images.   \label{tab:pos}} 
\begin{tabular}{cccccr}
&&&&& \\ \hline\hline
Image  & RA & Dec & RA & Dec & $S_{\rm 2.2~mm}$$^c$   \\
             & \multicolumn{2}{c}{ALMA mm-continuum$^a$} & \multicolumn{2}{c}{HST (CASTLES)$^b$} & \\
            & (\arcsec)      & (\arcsec) &(\arcsec) &(\arcsec) & $\mu$Jy  \\
\hline
A & 0             & 0                    & 0                                 & 0                                & 168 $\pm$ 31 \\
B &  $+0.050  \pm 0.005$  & $-0.192 \pm 0.005$             &  $+0.087 \pm 0.003$  &  $-0.163 \pm 0.003$ &  131 $\pm$ 31   \\
C & $+0.767 \pm 0.005$   & $-0.302 \pm 0.005$       &  $+0.774 \pm 0.003$  &  $-0.257 \pm 0.003$ &  27  $\pm$ 18   \\
D & $+0.643 \pm 0.005$   &  $+0.577  \pm 0.005$      &   $+0.610 \pm 0.003$ &  $+0.579 \pm 0.003$ &  41 $\pm$ 18  \\
X &  $+0.247 \pm 0.005$  &   $+0.247  \pm 0.005$     &                                     &                                   &  34 $\pm$ 18  \\
\hline 
\multicolumn{6}{l}{$^a$ The ALMA mm-continuum images A and B are extended. The ALMA RA and Dec positions}\\
\multicolumn{6}{l}{$^{~~~}$ correspond to the centroids the extended images.}\\
\multicolumn{6}{l}{$^b$ The HST image positions are taken from the CfA-Arizona Space Telescope LEns Survey}\\
\multicolumn{6}{l}{$^{~~~}$ (CASTLES) of gravitational lenses website http://cfa-www.harvard.edu/glensdata/.}\\
\multicolumn{6}{l}{$^c$ Flux densities of the ALMA images with the continuum integrated between 1.97mm and 2.17mm.}\\
\multicolumn{6}{l}{$^{~~~}$The fractional flux errors based on calibration are  $\sigma_{\rm F}$$/F$ $\sim$ 5\%.}\\
\end{tabular}
\end{table*}

\clearpage
\begin{table}
\caption{Results of lens modeling of 2~mm continuum and CO(J=3$\rightarrow$2) emission of  \hs\ \label{tab:lensmodel}} 
 \begin{tabular}{lcccrcc}
&&&&&&  \\ \hline\hline
Source$^a$ & $x_{\rm c}$  & $y_{\rm c}$   & $e$$^b$ & $\theta_{\rm e}$$^c$ & $\sigma_{\rm x}$$^d$ & $\sigma_{\rm y}$$^d$   \\ 
            &($\arcsec$)    & ($\arcsec$)   &        & ($^{\circ}$)         & ($\arcsec$)  & ($\arcsec$) \\
\hline
2~mm Continuum & $4$ $\times$ 10$^{-3}$  & $-8$ $\times$ 10$^{-3}$   & 0.97 & 65 & 0.079  & 0.018   \\
CO(J=3$\rightarrow$2)              &  $3$ $\times$ 10$^{-2}$  & $-2$  $\times$ 10$^{-2}$  & 0.69 & 110 & 0.047 & 0.034   \\
\hline 
\multicolumn{7}{l}{$^a$ The moment 0 images of the 2mm Continuum and CO(J=3$\rightarrow$2) line lensed}\\
\multicolumn{7}{l}{${~~~}$ emission observed in the high spatial resolution observation are modeled}\\
\multicolumn{7}{l}{$^{~~~}$ assuming extended sources having an elliptical gaussian geometry.}\\
\multicolumn{7}{l}{$^b$ The ellipticity of the source emission.}\\
\multicolumn{7}{l}{$^c$ The position angle of the source emission.}\\
\multicolumn{7}{l}{$^d$ $\sigma_{\rm x}$ and $\sigma_{\rm y}$ represent the widths of the gaussians along the major and minor}\\
\multicolumn{7}{l}{$^{~~~}$ axes, respectively. The FWHM values are 2.3548$\sigma$.}\\
\end{tabular}
\end{table}

\begin{table*}
\caption{Properties of redshifted and blueshifted clumps of CO(J=3$\rightarrow$2) emission \label{tab:clumps}} 
 \begin{tabular}{cccrcccrccc}
&&&&&&&&&& \\ \hline\hline
Clump$^a$ & RA & Dec & $v_{\rm offset}$ & FWHM & $F_{\rm d}$$^b$ &  $S/N$ & $\mu$$^c$ & $r_{\rm clump}$$^d$ &$M_{\rm clump}$ &  ${\dot{M}}_{\rm clump}$  \\ 
  &&&(km~s$^{-1}$) &(km~s$^{-1}$)  & (Jy~km~s$^{-1}$) & $\sigma$ &&kpc& $\times$~10$^{8}$~M$_{\odot}$& M$_{\odot}$/year\\
\hline
1 & 8:13:31.277  & $+$25:45:3.335 &  1294 $\pm$ 12 & 71 $\pm$  28    & $0.047 \pm 0.010$   & 4.7 & 18.3 & 0.95 $\pm$ 0.1 &  0.28 $\pm$ 0.07  & 39  $\pm$ 10\\
2 & 8:13:31.370  & $+$25:45:2.415 & 1141 $\pm$ 10     & 107 $\pm$ 22 & $0.101  \pm 0.033$  & 3.1& 1.7 & 10.9 $\pm$ 0.1 & 6.5 $\pm$  2.1  & 70 $\pm$ 23 \\
2 & 8:13:31.370  & $+$25:45:2.415 & $-1702$ $\pm$ 8 & 73 $\pm$ 18 & $0.074 \pm 0.023$     & 3.2 & 1.7 & 10.9 $\pm$ 0.1  & 4.7 $\pm$ 1.9  & 75 $\pm$ 34 \\
3 & 8:13:31.241  & $+$25:45:3.194 & $-890$ $\pm$ 6 & 64 $\pm$ 14   &$0.070 \pm 0.023$      & 3.1 &  2.2   &7.5 $\pm$ 0.1 & 3.5  $\pm$  1.5 & 43 $\pm$ 18\\
4 & 8:13:31.440    & $+$25:45:0.864 &  1018 $\pm$ 7   & 95 $\pm$ 17   &  $0.101  \pm 0.033$    &3.0 &  1.3  &   25.7  $\pm$ 0.1   & 8.5  $\pm$     2.4                    & 34 $\pm$ 10  \\  
5 & 8:13:31.424  & $+$25:45:2.107 & $-$891 $\pm$ 6   & 64 $\pm$ 13   &  $0.098  \pm 0.023$   &4.3 & 1.4  & 16.7 $\pm$ 0.1      &  7.6     $\pm$    1.8               & 41 $\pm$ 10\\ 
6 & 8:13:31.332  & $+$25:45:2.141 &  1267 $\pm$ 6   & 81 $\pm$ 15   &  $0.053  \pm 0.016$     & 3.3 & 1.8    &    9.9  $\pm$ 0.1     &  3.2    $\pm$   0.9               &  42 $\pm$ 12\\
7 & 8:13:31.273& $+$25:45:1.171 &  1304 $\pm$ 9   & 109 $\pm$ 20   &  $0.058  \pm 0.016$     & 3.5 &  1.4   &    18.3 $\pm$ 0.1    &   4.5      $\pm$ 1.2                &  33 $\pm$ 9\\ 
8 & 8:13:31.214& $+$25:45:3.503 &  $-$727 $\pm$ 7   & 78 $\pm$ 17   &  $0.042  \pm 0.011$     & 3.2 &  1.6  &   11.0 $\pm$ 0.1      &  2.9     $\pm$   0.8                &  20 $\pm$ 5\\ 
9 & 8:13:31.201& $+$25:45:0.852 &  $-$868 $\pm$ 7   & 73 $\pm$ 16   &  $0.024  \pm 0.006$    & 3.9 &  1.3  &   24.0 $\pm$ 0.1      & 2.0      $\pm$  0.6                  & 7 $\pm$ 2 \\ 
\hline
\multicolumn{11}{l}{$^a$ The clumps of CO(J=3$\rightarrow$2) emission are shown in Figure \ref{fig:alma_clumps}. Clump 2 shows both a redshifted and blueshifted component.}\\
\multicolumn{11}{l}{$^b$ Integrated flux densities are not corrected for magnification.}\\
\multicolumn{11}{l}{$^c$ The magnifications at the locations of the clumps were calculated from the magnification map of \hs\ described in Section 6.}\\
\multicolumn{11}{l}{$^d$ Clump 1 is located on the CO(J=3$\rightarrow$2) Einstein ring and based on our lens inversion is estimated to be at a distance of about 0.7~pc from the AGN.}\\
\multicolumn{11}{l}{For our assumed cosmology the scale at $z$ = 1.51 is 8.6kpc/\arcsec.}\\
\end{tabular}
\end{table*}




\bsp	
\label{lastpage}
\end{document}